\documentclass[journal]{IEEEtran}
\usepackage{cite}
\usepackage[pdftex]{graphicx}
\usepackage{epsf}
\usepackage{epstopdf}
\usepackage{graphicx}
\usepackage[cmex10]{amsmath}
\usepackage{eqnarray}
\usepackage[tight,footnotesize]{subfigure}
\usepackage{enumerate}
\usepackage{algorithm}
\usepackage{algorithmicx}
\usepackage{algpseudocode}
\usepackage[T1]{fontenc}
\usepackage{amsmath}
\usepackage{tikz}
\usepackage{multirow}

\begin{document}

		\title{Network Traffic Anomaly Detection}
		\author{\authorblockN{Hong Huang, Hussein Al-Azzawi, and Hajar Brani}
        \\
		\authorblockA{Klipsch School of Electrical and Computer Engineering\\New Mexico State University\\Las Cruces, NM, USA\\\{hhuang, azzawi, hajar\}@nmsu.edu.}}
	
	\maketitle
	\begin{abstract}
This paper presents a tutorial for network anomaly detection, focusing on non-signature-based approaches. Network traffic anomalies are unusual and significant changes in the traffic of a network. Networks play an important role in today's social and economic infrastructures. The security of the network becomes crucial, and network traffic anomaly detection constitutes an important part of network security. In this paper, we present three major approaches to non-signature-based network detection: PCA-based, sketch-based, and signal-analysis-based. In addition, we introduce a framework that subsumes the three approaches and a scheme for network anomaly extraction. We believe network anomaly detection will become more important in the future because of the increasing importance of network security.
	\end{abstract}

	\begin{IEEEkeywords}
		Network security, traffic anomaly, anomaly detection.
	\end{IEEEkeywords}

	\section{Introduction}
		\label{section:intro}
		Network traffic anomalies are unusual and significant changes in the traffic of a network. Examples of anomalies include both legitimate activities such as transient changes in the customer demand, flash crowds, etc., and illegitimate activities such as DDoS, port scans, virus and worms, etc. \cite{zhang2004}. Today, networks play an crucial role in our social and economic infrastructures. The security of the network becomes imperative, and network traffic anomaly detection constitutes an important part of network security. Despite its manifest importance, there is no good survey or tutorial papers on the subject of network anomaly detection. We hope this paper will remove this deficiency by providing a tutorial on network traffic anomaly detection. We believe network anomaly detection will become more important in the future and the knowledge about it will become more useful.

There are two major approaches to network anomaly detection: signature-based and non-signature-based. In the signature-based approach, anomaly is detected by looking for patterns that match signatures of known anomalies. For example, DoS activities can be discovered based on the uniformity of IP addresses \cite{moore2001}. The limitation of this approach is that it requires the signature to be known beforehand, and it is not capable to detect new anomalies. In the second approach, statistical techniques are applied to network traffic. This approach does not require any prior knowledge about the anomalies, and it is capable of discovering new anomalies. This paper focuses on non-signature-based anomaly detection.

In this paper, we describe three major approaches to non-signature-based anomaly detection: principal-component-analysis (PCA)-based, sketch-based, and signal-analysis-based. Comparison among approaches is difficult because of three reasons: 1) There are many types of anomalies. Evaluation studies typically focus on only a subset of anomalies, and different studies use different data sets and focus on different subsets of anomalies. 2) No existing method is consistently better than the others in dealing with different types of anomalies \cite{nyalkalkar2011}. 3) The field is still evolving and further optimizations are still emerging in each approach. Generally speaking, the sketch-based approach requires less computational complexity and storage capacity than the other two approaches, but the other two approaches sometimes achieve better detection performance. In section V, we provide a framework that synthesizes the three approaches.

\begin{table*}[!ht]
\caption {Various Approaches to Network Anomaly Detection}
\begin{center}
\begin{tabular}{ l|l|l|l }
 \hline
 Approaches & Progression of schemes & Sections in the paper & References \\ \hline
 \multirow{5}{*}{PCA-based} & The basic scheme & III.A-C & \cite{lakhina2004} \\
  & Using traffic feature distributions & III.D & \cite{lakhina2005} \\
  & Using sketch subspaces & III.E & \cite{li2006}\\
  & Detecting small-volume, correlated anomalies& III.F &  \cite{silveira2010} \\
  & Distributed PCA & III.G & \cite{huang2007}\\
  & Problems and solutions with the PCA-based approach & III.H &  \cite{ringberg2007},\cite{brauckhoff2009b} \\ \hline
 \multirow{3}{*}{Sketch-based} & The basic scheme & IV.A-C & \cite{muthukrishnan2003} \\
  & Identifying hierarchical heavy hitters & IV.D & \cite{zhang2004} \\
  & Using sketches and non-Gaussian multi-resolution statistics & IV.E & \cite{dewaele2007} \\ \hline
 \multirow{3}{*}{Signal-analysis-based} & Statistics-based & V.A & \cite{thottan2003} \\
  & Wavelet-based & V.B & \cite{barford2002} \\
  & Kalman-filter-based & V.C &\cite{soule2005} \\ \hline
  Synthesis of different approaches & Network anomography & VI.A & \cite{zhang2005} \\ \hline
  Anomaly extraction & Using associate rule mining & VI.B & \cite{brauckhoff2009} \\ \hline

 \end{tabular}
 \end{center}
\label{table:sum_approaches}
\end{table*}

Today's network anomaly detection schemes have evolved to highly sophisticated levels, involving advanced signal processing techniques such as PCA, sketches, time series analysis, wavelets, Kalman filter, etc. In order for this tutorial to be useful to readers with different backgrounds, we present the material at two levels. At the first level, which consists of Section II, we present the basic concepts of three representative network anomaly detection approaches: PCA-based, sketch-based, and wavelet-based. The first level lays out the basic ideas underlying the main approaches to network anomaly detection, and it does not require much background in signal processing and thus is suitable for the nonexpert.

The remainder of the paper constitute the second level, which is intended for the expert. The reader should have strong background in signal processing and be prepared to be exposed to highly advanced signal processing techniques, the descriptions of which involve a fair amount of mathematical formulas. Since the purpose of the tutorial is to enable the advanced reader to implement their own network anomaly detection schemes, we provide very detailed description for the techniques covered in the paper. For each of three network anomaly detection approaches, we describe a progression of anomaly detection schemes, from the basic to the more refined, except for the third approach where the schemes described there are independent of each other. See Table I for an overview of the network anomaly detection schemes described in the paper.


The paper is organized as follows. In Section II, we introduce the basic concepts in network anomaly detection. Sections III, IV, and V, we introduce three approaches to non-signature-based anomaly detection: PCA-based, sketch-based, and signal-analysis-based. In Section VI, we describe a framework that subsumes the three approaches and also a scheme for anomaly extraction. We conclude in Section VII.


     \section{Basic Concepts}
		\label{section:basic}
		In the following, we first define network traffic anomaly and then introduce the basic ideas of three representative network anomaly detection approaches: PCA-based, sketch-based, and wavelet-based.
\subsection{Network Traffic Anomaly}
Network traffic anomalies are unusual and significant changes in the traffic of a network. These anomalies can be changes in link traffic volume, distribution patterns of IP source and/or destination addresses and port numbers, etc. The causes of anomalies include both legitimate and illegitimate activities \cite{zhang2004}. Legitimate activities include transient changes in the customer demand, flash crowds, routing table changes, etc. Illegitimate activities include DDoS, port scans, virus and worms, etc.
\begin{figure*}[ht]
\begin{center}
\includegraphics[width=7in]{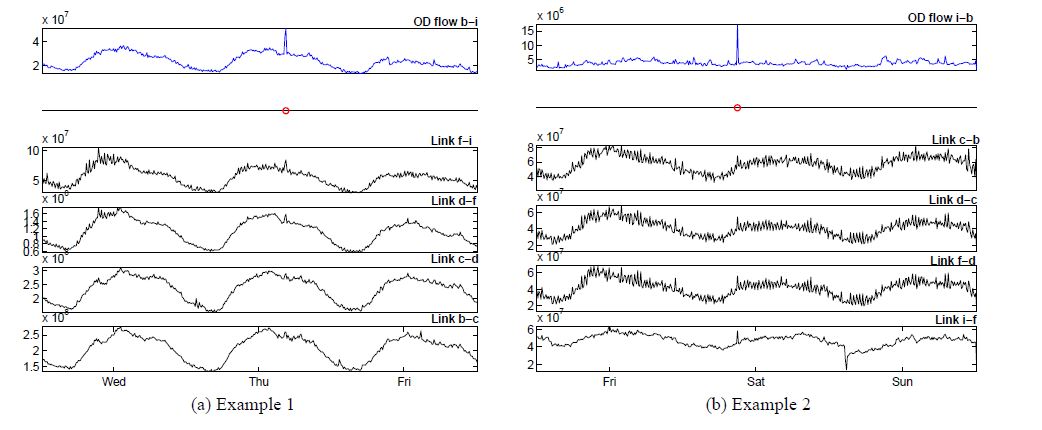}
\end{center}
\caption{Illustration of anomalies at the OD flow level (top row) and at link traffic level (lower four rows) \cite{lakhina2004}.} \label{vol-anomaly}
\end{figure*}
In Figure \ref{vol-anomaly}, we show two examples of anomalies of traffic volume time series data both at the origin-designation (OD) flow level and at the link traffic level \cite{lakhina2004}. Anomaly at OD flow level, which is not directly measurable, causes anomaly at the link traffic level, which is measurable. The anomalies at the OD flow level, occurring at the time instant designated by the red dots, are quite pronounced upon visual inspection. However, anomalies at the link traffic level are much less pronounced visually. For example, the anomalies on links c-d and b-c in Example 1 are hardly discernible. Therefore, network anomaly detection is a challenging problem.

\subsection{The PCA-Based Approach}
PCA is a coordinate transformation that maps a set of $n$-dimensional data points onto $n$ new axes called principal axes $v_i, i =1,2,..n$. The principal components have the following properties. The first principal component points in the direction of maximum variance of the data. The second principal component points in the direction of maximum variance remaining in the residue data after removing variance already accounted by the first principal component, and so on. Therefore, the principal components are ordered by the amount of variance they account for. PCA is commonly used for dimension reduction. If most of the variance of $n$-dimensional data is accounted by $k < n$ principal components, then the dimension of the data can be reduced to $k$. In the following, we describe how to apply PCA to network anomaly detection, first based on traffic volumes and then based on traffic features such as IP addresses and port numbers.

\subsubsection{Volume-Based Detection}
PCA can be used as an effective technique for detecting traffic volume anomalies as follows \cite{lakhina2004}. The input of PCA is the $m\times n$ traffic matrix $A$, where $a_{i,j}$ is the traffic volume on the $j$-th link at the $i$-th time interval, $m$ is the number of time slots in the measurement window, and $n$ is the number of links in the network. We normalize $A$ to obtain the $m\times n$ matrix $X$, where $x_{i,j}$ has zero mean and unit variance. We apply PCA to $X$. Let $u_i$ denote the projection of data onto the principal axis $i$. Thus, $u_1$ captures the most of the variance of the data, $u_2$ the second most of the variance, and so on. We set a certain empirical threshold $u_T$ such that all the $u_i \geq u_T$ belong to the normal set and all the $u_i < u_T$ belong to the abnormal set. The corresponding principal axes form the normal set of axes $V_1$ and the abnormal set of axes $V_2$. We project the link traffic vector $x$ (a certain column of $X$) onto $V_1$ and $V_2$ to obtain $x_1$ and $x_2$. We declare an anomaly if $\|x_2\|^2$ is larger than a threshold. The threshold is determined by the required confidence level using the statistical test called Q-Statistics \cite{jackson1979}.

\begin{figure}[ht!]
\begin{center}
\includegraphics[width=3.5 in]{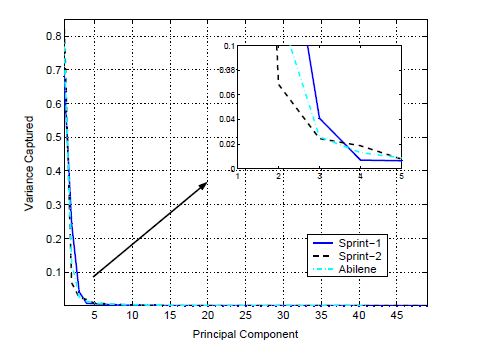}
\end{center}
\caption{Link traffic variance captured by principal components\cite{lakhina2004}.} \label{low-dimension}
\end{figure}

The effectiveness of PCA in network anomaly detection can be explained by the fact that the link traffic has low effective dimension. Figure \ref{low-dimension} shows the link traffic variance captured by principal components in three network scenarios. We can see that the first 3 or 4 principal components capture most of the variance. This low effective dimensionality of traffic data forms the basis for the effectiveness of PCA in network anomaly detection.

\begin{table*}[!ht]
\caption {Traffic feature distributions affected by various anomalies}
\begin{center}
\begin{tabular}{ p{3cm}  p{8cm} p{5cm} }
\hline
Anomaly  & Definition   &Traffic feature distributions affected  \\ \hline
Alpha flows     &Large-volume point-to-point flows  &SIP, DIP (possibly SP, DP)                         \\
DoS              &Denial of service attack (distributed or single-source)  &SIP, DIP                         \\
Flash crowd     &Large volume of traffic to a single destination, typically from a large number of sources  &DIP DP                         \\
Port scan     &Probe to many destination ports on a small number of destination addresses  &DIP, DP                         \\
Network scan     &Probe to many destination addresses on a small number of destination ports &DIP, DP                         \\
Outage events     &Traffic shifts because of equipment failures or maintenance  &SIP, DIP                         \\
Point-to-multipoint    &Traffic from a single source to many destinations, e.g., content distribution  &SIP, DIP                         \\
Worms     &Scanning by worms for vulnerable hosts, which is a special case of network scan  &DIP, DP                         \\
\\ \hline
\end{tabular}
\end{center}
\label{table:pca_feature}
\end{table*}

\begin{figure*}[ht]
\begin{center}
\includegraphics[width=7.5 in]{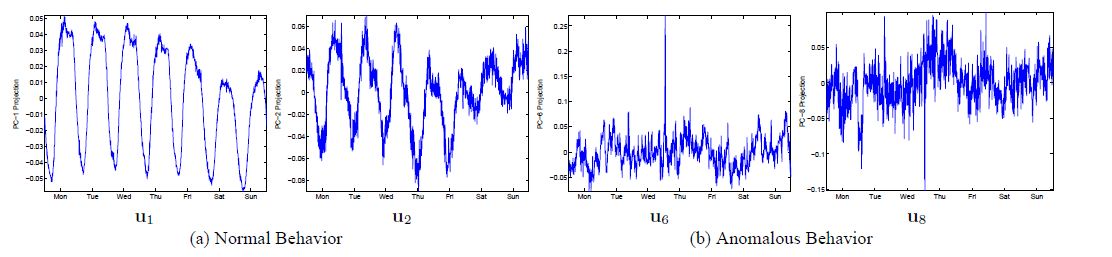}
\end{center}
\caption{Traffic projections onto the normal and anomalous principal axes \cite{lakhina2004}.} \label{normal-abnormal}
\end{figure*}

Figure \ref{normal-abnormal} shows traffic projections onto the normal principal axes ($u_1, u_2$) and anomalous principal axes ($u_6, u_8$). The normal projections on the left capture the most variance in the data. These time series data are quite regular and roughly periodic, and they reflect the typical diurnal traffic patterns. In contrast, the anomalous projects on the right exhibit abrupt "spikes" indicative of traffic anomalies.

\subsubsection{Feature-Based Detection}
\begin{figure}[ht!]
\begin{center}
\includegraphics[width=3.5 in]{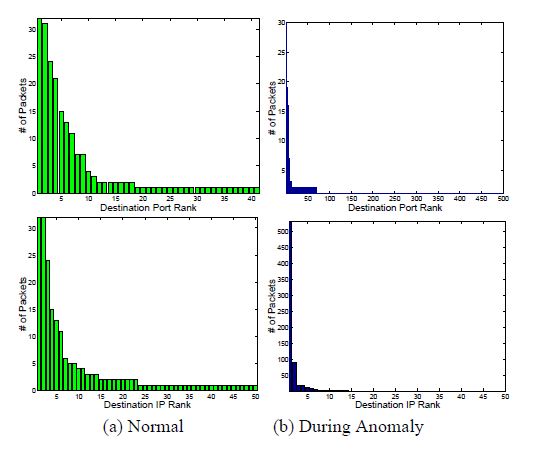}
\end{center}
\caption{Distribution changes caused by a port scan anomaly. Upper and lower left: normal DP and DIP distributions. Upper right: anomalous DP distribution is dispersed (notice the x-scale is expanded). Lower right: anomalous DIP distribution is concentrated (notice that the y-scale is expanded) \cite{lakhina2005}.} \label{traf-feature}
\end{figure}

PCA-based network anomaly detection can be extended beyond traffic volumes to other traffic features such as source and destination IP addresses/port numbers \cite{lakhina2005}. In Table \ref{table:pca_feature}, traffic feature distributions affected by various anomalies are listed. Anomalies cause the dispersal or concentration of traffic feature distributions. For example, flows that have large volumes, the so-called Alpha flows, will cause source and destination IP addresses (SIP, DIP), and possible port numbers (SP, DP), concentrated on a few values associated with the Alpha flows. A DoS attack will cause a concentration of SIP and DIP on those of attackers and victims. A flash crowd will cause a concentration of DIP and DP on those of the flash target. A port scan will cause a dispersed distribution on DP but a concentrated distribution of DIP, as shown in Figure \ref{traf-feature}. In contrary, a network scan will cause a dispersed distribution of DIP but a concentrated distribution of DP, and so on.

\begin{figure}[ht!]
\begin{center}
\includegraphics[width=3.5 in]{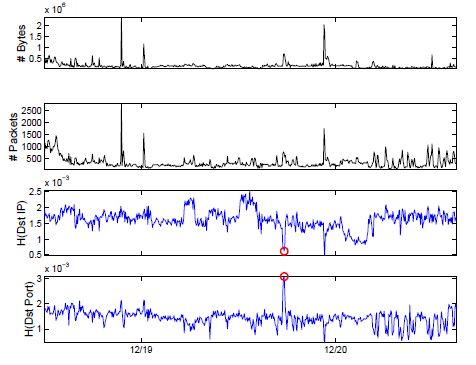}
\end{center}
\caption{Port scan anomaly. First and second rows: traffic volume distributions in bytes and in packets, respectively. Third and fourth rows: DIP and DP entropy distributions, respectively \cite{lakhina2005}.} \label{traf-feature2}
\end{figure}

The procedure of applying PCA to various traffic features is very similar to that of applying PCA to traffic volumes. The only difference is that, instead of the traffic volume matrix $X$, we apply PCA to an traffic feature entropy matrix $H$, whose element $h_{i,j}$ denote the entropy of the $j$-th traffic feature distribution at time instant $i$. Traffic feature entropy provides an effective metric for capturing the dispersal or concentration of traffic feature distributions. For details, the reader is referred to Section III.D.

Compared with volume-based anomaly detection, feature-based anomaly detection has two advantages. First, it can detect anomalies with minor volume changes, such as scans or small DoS attacks. Second, changes in feature distribution provides useful information about the structure of anomalies and can be used in the classification of anomalies. Figure \ref{traf-feature2} shows the effectiveness of feature-based anomaly detection in the port scan anomaly occurring at the time indicated by the red circle. The first two rows of the figure shows the traffic volume distributions in bytes and packets respectively. The anomaly is hardly detectable on the basis of traffic volume. However, the anomaly is easily detectable on the basis of traffic feature entropy. Since port scan concentrates on a single DIP and is dispersed on DP so that the DIP entropy is low and the DP entropy is high when the anomaly occurs. Finally, it was shown in \cite{lakhina2005} that the anomalies detected using volume-based and feature-based detections are largely disjoint, with the former detecting anomalies that have large impact on traffic volumes and the latter detecting those that have impact on traffic feature distributions. Thus, volume-based and feature-based anomaly detection methods are complimentary to each other.

\subsection{The Sketch-Based Approach}
The method of change detection \cite{arsham} can be used to detect network anomaly. It works by deriving a model of normal traffic behavior based on past traffic history and searching for significant changes in observed behavior that deviates from the model. The standard modeling techniques include smoothing such as sliding window averaging, exponential smoothing, and the Box-Jenkins ARIMA modeling \cite{box1976, box1994}, the details of which are given in Section IV.B. However, the traditional change detection techniques are not scalable to a large number of time series data typically seen in network anomaly detection. For example, if we apply change detection on a per-flow basis, then the total number of all possible flows is $2^{(32+16)\times 2} = 2^{104}$, since each flow is defined by 32-bit source and destination IP addresses and 16-bit source and destination port numbers. Obviously, we have a scalability problem.

The solution to the scalability problem is to use data stream computation, which is an effective technique to process massive data streams online \cite{muthukrishnan2003}. Using this technique, data is processed exactly once. One particular technique of data stream computation is the sketch, which is a probabilistic summary method that uses random projections. Let $I = I_1,I_2,...$ denote an input stream that arrives sequentially. Each item $I_i = (a_i, u_i)$ is composed of a key $a_i \in \{0, 1,..., u - 1\}$, and an update $u_i \in R$. Associated with each key $a_i$ is a signal $A[a_i]$. The arrival of item $I_i$ causes an update as follows
        \begin{equation}
		\label{eqn:bc_sketch_update1}
			A[a_i] = A[a_i] + u_i.
		\end{equation}
Compared to traditional data structures, sketches have the following advantages. They are space efficient and provide reconstruction accuracy guarantees \cite{muthukrishnan2003}.

\begin{figure}[ht!]
\begin{center}
\includegraphics[width=3.5 in]{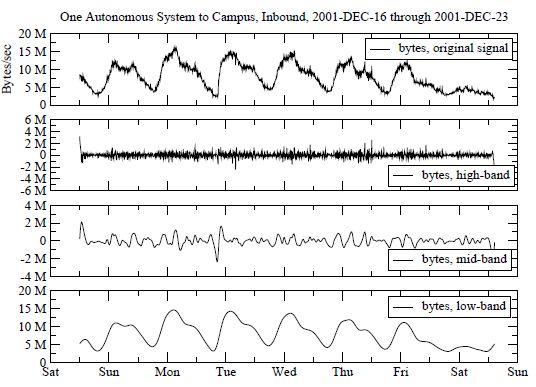}
\end{center}
\caption{Aggregate traffic bytes for a week with no anomaly and the high/mid/low bands \cite{barford2002}.} \label{wav-normal}
\end{figure}

\begin{figure}[ht!]
\begin{center}
\includegraphics[width=3.5 in]{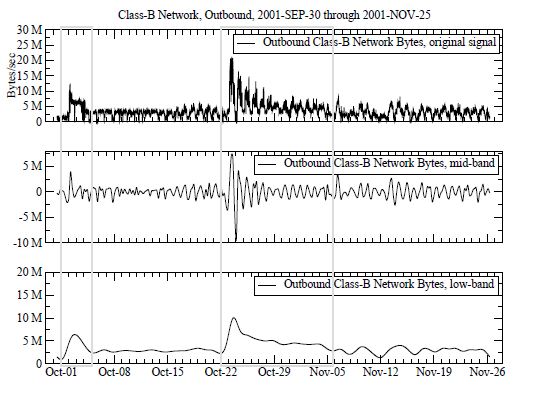}
\end{center}
\caption{Aggregate traffic bytes for a week with flash crowds (two grey boxes) and the mid/low bands \cite{barford2002}.} \label{wav-flash}
\end{figure}

\begin{figure}[ht!]
\begin{center}
\includegraphics[width=3.5 in]{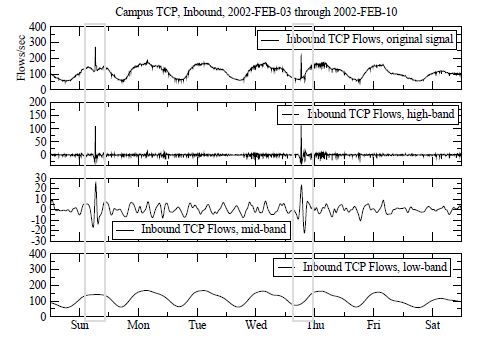}
\end{center}
\caption{Aggregate flow counts for a week with DoS (two grey boxes) and the high/mid/low bands \cite{barford2002}.} \label{wav-dos}
\end{figure}

In the context of network anomaly detection, we can use one or more fields in the IP header as the key. The update can be the packet size, the total number of bytes or packets in a traffic flow, etc. 
The sketch-based network anomaly detection consists of three steps \cite{krishnamurthy2003}. First, we create a number of sketches to summarize the traffic behavior. Second, we build various forecasting models on top of the sketches. Third, we compare the observed sketch with the forecast sketch and declare an anomaly if the difference exceeds a certain threshold. This sketch-based anomaly detection scheme requires a constant, small amount of memory, and has constant update cost. The reader is referred to Section IV for more details.

\subsection{The Wavelet-Based Approach}
A wavelet-based approach was proposed in \cite{barford2002} for network anomaly detection. The wavelets provide a powerful combined time-frequency characterization of the signal. The traffic stream, sampled every 5 min, is treated as a generic signal. The wavelet analysis decomposes the traffic into bands. The lower bands contain low-frequency or aggregated information and can be used to detect long-lived anomalies such as flash crowd events that can last up to a week. In contrast, the higher bands contain high-frequency or fine-grained information and can be used to detect short-lived anomalies such as network attack, failures, etc. An anomaly test is developed in \cite{barford2002} that is based on the local variance of the mid and high frequency bands of the signal. An anomaly is declared if the local variance exceeds a certain threshold, indicating unpredictable changes in traffic behavior.

In the following, we show some examples of using wavelets for anomaly detection. Figure \ref{wav-normal} shows the wavelet decomposition of the traffic volume signal into high/mid/low bands. The weekly cycle of traffic variation is clearly visible in the low band. Figure \ref{wav-flash} shows the wavelet decomposition of traffic signal into mid/low bands when there are flash crowds (long-lived heavy traffic anomaly). The anomaly is clearly visible in the low-band due to the long-lived nature of the anomaly. Figure \ref{wav-dos} shows the wavelet decomposition of traffic signal into high/mid/low bands when there are DoS attack (short-lived heavy traffic anomaly). In contrast to flash crowds, the anomaly is clearly visible in the high/mid bands due to the short-lived nature of the anormaly. For more details, the reader is referred to Section V.B.  
	\section{PCA-Based Approaches}
		\label{section:pca}
		In this section, we first describe a basic anomaly detection scheme using PCA in subsections A, B, and C. Then, we introduce more refined PCA-based schemes using traffic feature distributions and sketch subspaces in subsections D and E. In subsection F, we introduce a scheme that specializes in detecting small-volume, correlated anomalies instead of large-volume anomalies that other detection schemes specialize in. In subsection G, we introduce a communication-efficient scheme that detects anomaly in a distributed fashion. We end the section by discussing potential problems with the PCA-based approach and the solutions to the problems.

\subsection{Anomaly Detection}
A method based on PCA was proposed to diagnosing network-wide traffic volume anomalies in \cite{lakhina2004}, which proved to be highly effective. This method is in contrast to much of the prior work in anomaly detection, which focuses on single-link traffic data. The network-wide view enables the detection of anomalies that may be too small in the individual link to be detectable by a single-link detector. The method separates traffic into normal and abnormal subspaces using PCA. The method can simultaneously achieve three objectives: 1) detecting anomalies; 2) identifying the underlying origin-destination (OD) flows that are the sources of the anomalies; and 3) estimating the amount of traffic involved in the anomalies. In the following, we first introduce PCA and then describe how the three objectives are achieved.

\subsubsection{PCA}
PCA is a coordinate transformation that maps a set of data points onto new axes called principal axes. PCA requires that the input data has zero mean and unit variance. The principal components have the following properties. The first principal component points in the direction of maximum variance. The second principal component points in the direction of maximum variance remaining in the residue data after removing variance already accounted by the first principal component, and so on. Therefore, the principal components are ordered by the amount of variance they account for. PCA can be used for dimension reduction. If most of the variance of $n$-dimensional data is accounted by $k < n$ principal components, then the dimension of the data can be reduced to $k$.

Let $A$ denote the $m\times n$ traffic measurement matrix, where its element $a_{i,j}$ is the traffic volume on the $j$-th link at the $i$-th time interval. Let $X$ be the normalized version of $A$, i.e.,
        \begin{equation}
		\label{eqn:pca_x}
			x_{i,j} = a_{i,j} - \bar{a}_{j}, \quad\text{and}\quad \bar{a}_{j} = \frac{1}{m}\sum_{i = t - m + 1}^t{x_{i,j}}.
		\end{equation}

We apply PCA on $X$, which results in $n$ principal components. The first principal component is the vector that points in the direction of maximum variance in $X$ and is given by
        \begin{equation}
		\label{eqn:pca_v_1}
			v_1 = \underset{\|v\| = 1}{\text{arg min}} \|Xv\|
		\end{equation}
where $\|v\|$ is the Euclidean norm of $v$, and $\|Xv\|$ is proportional to the data variance measured along the vector $v$. Given the first $i - 1$ principal components $v_1,v_2,...v_{i-1}$, the $i$-th principal component is given by
        \begin{equation}
		\label{eqn:pca_v_i}
			v_i = \underset{\|v\| = 1}{\text{arg min}} \|X(1 - \sum_{j = 1}^{i-1}v_jv^T_j) v\|
		\end{equation}
where the $i$-th principal component captures the maximum variance in the residue after those of first $i - 1$ have been removed. The $v_i$'s are also the eigenvectors of the covariance matrix $C$ of $X$ given by
        \begin{equation}
		\label{eqn:pca_cov}
			C = \frac{1}{m} X^T X.
		\end{equation}

After the principal axes have been found, we can map the data set onto the new axes as follows
        \begin{equation}
		\label{eqn:pca_u}
			u_i = \frac{Xv_i}{\|Xv_i\|}\quad i \in{[1,n]}
		\end{equation}
where $u_i$ is an $m$-dimensional vector, and represents the projection of data onto the principal axis $i$. Thus, $u_1$ captures the most of the variance of the data, $u_2$ the second most of the variance, and so on.

\subsubsection{Subspace Method}
The subspace method works by separating the principal axes into two sets: normal set $S_1$ and abnormal set $S_2$. A threshold-based procedure can be used for the separation. Specifically, we examine the data projection on each axis sequentially, from $u_1$ to $u_n$. We stop as soon as the data projection $u_{k+1}$ crosses the threshold, i.e., the projection has a $3\sigma$ deviation from the mean. Then, the first $k$ principal components are classified as residing in the normal subspace, and the last $n - k$ principal components in the abnormal subspace.

We decompose link measurements $x$ into normal part $x_1$ and abnormal part $x_2$
        \begin{equation}
		\label{eqn:pca_x_decomp}
			x = x_1 + x_2.
		\end{equation}
The parts $x_1$ and $x_2$ are given by projections onto the normal and abnormal principal axes $(v_1,v_2,...v_k)$ and $(v_{k+1}, v_{k+2},...v_n)$, respectively, i.e.,
        \begin{equation}
		\label{eqn:pca_x_proj}
			x_1 = Px \quad\text{and}\quad x_2 = (1 - P)x
		\end{equation}
where the projection matrix is given by
        \begin{equation}
		\label{eqn:pca_u}
			P = V_k V^T_k
		\end{equation}
where $V_k$ is an $m\times k$ matrix composed of $(v_1,v_2,...v_k)$.

We define squared prediction error SPE as
        \begin{equation}
		\label{eqn:pca_spe}
			\text{SPE} = \|x_2\|^2 = \|(1 - P)x\|^2
		\end{equation}
and we classify network traffic as normal if
        \begin{equation}
		\label{eqn:pca_traf_cret}
			\text{SPE} \leq \sigma_{\alpha}^2
		\end{equation}
where $\sigma_{\alpha}^2$ is the threshold for the SPE at the $1 - \alpha$ confidence level.

A statistical test called Q-Statistics is given below \cite{jackson1979}
        \begin{equation}
		\label{eqn:pca_q_test}
			\sigma_{\alpha}^2 = \phi_1(\frac{c_\alpha \sqrt{2\phi_2h_0^2}}{\phi_1} + \frac{\phi_2 h_0(h_0 - 1)}{\phi_1^2} + 1)^{1/h_0}
		\end{equation}
where
        \begin{equation}
		\label{eqn:pca_q_test}
			h_0 = 1 - \frac{2\phi_1\phi_3}{3\phi_2^2}, \quad\text{and}\quad \phi_i = \sum_{j = k + 1}^{n}\lambda_j^i, i = 1, 2, 3
		\end{equation}
where $\lambda_j = \|Xv_j\|$ is the variance captured by projecting data on the $j$-th principal axis, $c_\alpha$ is the $1 - \alpha$ normal quantile, with $\alpha$ being the false alarm probability.

\subsection{Anomaly Identification}
We assume the set of all possible anomalies is $\{F_i, i = 1,2,...I\}$. For the ease of exposition, we consider one-dimensional anomalies, and the generalization to multi-dimensional anomalies is straight forward. We associate each anomaly $F_i$ with an associated unit-norm vector $\theta_i$, which defines how the anomaly adds traffic to the network. With the anomaly, the link traffic vector is modified to
        \begin{equation}
		\label{eqn:pca_x_anom}
			x = x_1 + \theta_i f_i
		\end{equation}
where $x_1$ is the normal traffic, and $f_i$ is the magnitude of the anomaly.

We can estimate $f_i$ by minimizing the distance to the abnormal subspace in the direction of the anomaly as follows
        \begin{equation}
		\label{eqn:pca_fi_estimate}
		\hat{f_i} = \underset{f_i}{\text{arg min}}\|x_2 - \theta_{i,2}f_i\|	
		\end{equation}
where $x_2$ is the abnormal traffic, and $\theta_{i,2} = (1 - P)\theta_i$. As a result of the minimization, we have
        \begin{equation}
		\label{eqn:pca_fi_value}
			\hat{f_i} = (\theta_{i,2}^T\theta_{i,2})^{-1}\theta_{i,2}^Tx_2.
		\end{equation}
Given $F_i$, the best estimate for $x_{1,i}$ is given by
        \begin{eqnarray}
        \label{eqn:pca_x1_estimate}
            \hat{x}_{1,i}&=& x - \theta_i\hat{f_i}      \nonumber \\
                         &=& x - \theta_i(\theta_{i,2}^T\theta_{i,2})^{-1}\theta_{i,2}^Tx_2 \nonumber \\
                         &=& (I - \theta_i(\theta_{i,2}^T\theta_{i,2})^{-1}\theta_{i,2}^T(1 - P))x.
        \end{eqnarray}

The identification algorithm is as follows:
\begin{itemize}
\item  For each anomaly $F_i$, compute $\hat{x}_{1,i}$ according to (\ref{eqn:pca_x1_estimate}).
\item  Choose anomaly $F_j$ as follows
        \begin{equation}
		\label{eqn:pca_j_select}
			j =\text{arg }\underset{i}{\text{min}}\|(1 - P)\hat{x}_{1,i}\|.
		\end{equation}
\end{itemize}
In \cite{lakhina2004}, only the anomalies coming from a single OD flow were considered. Thus, the set of anomalies are $\{F_i, i = 1,2,...n\}$, each corresponding to a particular flow.
\subsection{Anomaly Quantification}
The estimated amount in bytes of anomalous traffic contributed by anomaly $F_i$ is given by
        \begin{equation}
		\label{eqn:pca_y2_amount}
			x_2 = x - \hat{x}_{1,i}.
		\end{equation}

For an anomaly to be detectable, it can not lie completely in the normal subspace. That is, if $(1 - P)\theta_i = 0$, then anomaly $F_i$ is not detectable. A sufficient condition to successfully detect anomaly is given by \cite{dunia1979}
        \begin{equation}
		\label{eqn:pca_condi_detect}
			f_i > \frac{2\delta_\alpha}{\|(1 - P)\theta_i\|}
		\end{equation}
which basically says that the larger the projection of the normalized anomaly onto the abnormal subspace, the easier it is to detect.

Finally, we note that computational complexity of PCA is $O(mn^2)$\cite{golub1996}, which makes it feasible for online detection. In \cite{lakhina2004}, it was shown that PCA-based diagnostic scheme out-performs significantly other diagnostic schemes, such as the exponential weighted moving average (EWMA) \cite{brutlag2000,krishnamurthy2003}, and those using signal processing techniques (wavelet/Fourier analysis)\cite{barford2002}. The PCA-based scheme can also be used for other metrics, such as the number of flows and the size of packets in the network.
\subsection{Anomaly Detection Using Traffic Feature Distributions}
In \cite{lakhina2005}, the subspace method is extended to detect anomalies using traffic features, such as source and destination IP addresses (SIP, DIP), and source and destination port numbers (SP, DP). This brings two benefits: 1) It can detect a wide range of anomalies with high sensitivity, augmenting that of volume-based method. 2) It enables automatic anomaly classification based on unsupervised learning. Anomaly classification has not been satisfactorily addressed until then, especially when the anomalies do not cause detectable volume changes. Since anomaly detectors proposed previously in the literature were volume-based, no detectable volume changes meant that the anomalies were not detectable. In the following, we first describe the extension of the subspace method to the multi-way subspace method, which is applied to multiple traffic feature distributions, and then we describe anomaly classification.

\subsubsection{The Multi-way Subspace Method}
Anomalies cause the dispersal or concentration of traffic feature distributions , which can be captured by the sample entropy. Given an empirical histogram $X = (n_i, i = 1, 2, ...N)$, where $n_i$ indicates that traffic feature $i$ occurred $n_i$ times in the sample, the sample entropy is defined as
        \begin{equation}
		\label{eqn:pca_entropy}
			H(x) = - \sum_{i=1}^N (\frac{n_i}{S})\log_2(\frac{n_i}{S})
		\end{equation}
where $S = \sum_{i=1}^N n_i$ is the sample size. The value of sample entropy is in the range $[0, \log_2N]$. The sample entropy is 0 when the distribution is maximally concentrated, i.e., there is only one feature present. The sample entropy is $\log_2N$ when the distribution is maximally dispersed, i.e, all features appear with the same frequency.

The multi-way subspace method is used to enable anomaly detection across multiple features simultaneously and across multiple flows in the network. For example, a threeway sample entropy matrix consists elements $H(t,p,k)$, which presents the entropy at time $t$, for flow $p$, and of traffic feature $k$. Let $H(SIP), H(DIP), H(SP), H(DP)$ denote the entropy matrices for source and destination IP addresses/port numbers.

Using a standard technique in multivariate statistics, the multi-way data is recast to a single-way representation. Since we are interested in four features, we recast $H(t,p,4)$ to $H(t,4p)$, where we abused the notation a little by letting $t, p$ denote sizes of the dimensions. That is, the first $p$ columns of $H(t, 4p)$ represent SIP entropy, the next $p$ columns DIP entropy, the next $p$ columns SP entropy, and the last $p$ columns DP entropy.

Next, we can apply standard subspace method to $H(t,4p)$ by decomposing it into normal part $H_1$ and subnormal part $H_2$ at time $t$ as below
        \begin{equation}
		\label{eqn:pca_sub_h}
			H = H_1 + H_2.
		\end{equation}
Anomalies can be detected by testing $|H_2|$ against a threshold, which can be determined by the desired false alarm rate, as mentioned in the earlier subsection. Also, we can adapt single-way subspace method for anomaly identification to multi-way data \cite{lakhina2005}.
\subsubsection{Unsupervised Classification}
Unsupervised classification is used, since it can adapt to new anomalies. Specifically, a clustering approach is used. There are two types of clustering algorithms: partitional and hierarchical. Partitional algorithms take a top-down approach and divide the global data into $k$ clusters. Hierarchical algorithms take a bottom-up approach and meager neighbors into clusters, and smaller clusters into larger clusters. We use two representative algorithms from each type. Namely, the $k$-means algorithm from partitional algorithms, and the hierarchical agglomerative algorithm from hierarchical algorithms. The distance metric used is Euclidean distance. It was shown that the results are insensitive to the choice of algorithms \cite{lakhina2005}. The following method is used to select the proper number of clusters. We increase the number of clusters to the extent that intra-cluster variations reach a minimum and inter-cluster variations reach a maximum. Thus, adding additional cluster has marginal benefits.

Evaluations form sample data collected from two tier-1 backbone networks show that the feature-distribution-based method provide better performance than traditional methods such as the volume-based ones, especially when the anomalies have low traffic volumes. One of the reasons that the multi-way method is more effective is that some low-volume anomalies exhibit strong simultaneous changes across multiple traffic features, which makes the multi-way approach effective. For example, a port scan exhibits simultaneously a dispersal in destination ports and a concentration in destination IP addresses.

Before moving on to the next subsection, we mention four related works that also use traffic feature distributions. In \cite{lee2001}, entropy and conditional entropy are used to provide data partitioning and parameter setting for intrusion detection. In \cite{gu2005}, an anomaly detection scheme was proposed, which works by comparing the current network traffic with a baseline distribution. The maximum entropy technique provides a fast and flexible approach for estimating the baseline distribution. The anomaly detection scheme consists of two phases. In the first phase, the baseline distribution is learned. In the second phase, the anomaly detection is performed. In \cite{kind2009}, the authors extend the anomaly detection method using histograms of traffic features. They investigate the utility of different features, the construction of feature histograms, the modeling and clustering algorithms, and the detection of deviations. Compared to previous feature-based methods, their approach constructs detailed histogram models, rather than using the coarse approximation of the entropy-distributions. In \cite{nychis2008}, empirical evaluations of entropy-based anomaly detection is performed. Two classes of distributions are considerd: 1) flow-head features: IP addresses, port numbers, and flow sizes, 2) behavioral features: the number of IP addresses each host communicate with. The evaluations show that entropy values of IP address and port distributions are strongly correlated and provide very similar anomaly detection capabilities. Those of behavioral and flow size distributions are less correlated and often provide better anomaly detection capabilities.
\subsection{Sketch Subspaces}
In \cite{li2006}, the methods based on subspaces and sketches are combined to detect anomalies with high accuracy. Sketch is a particular technique of data stream computation, where data is processed exactly once, and which will described in detail in the next section. The hybrid method, called $Defeat$, can also be used for identification, whereas the sketch-only-based method can not. $Defeat$ is based on the insight that the global traffic sketches preserve the normal traffic variation and most of the residual subspace. In this method, multiple sketches of feature entropies of the global traffic are taken, to which the subspace method is applied. Because the anomalies are shuffled among different sketches, agreement among sketches can be used to detect anomalies robustly.

We consider a network with $n$ routers $R_1, R_2,...,R_n$. Measurements are performed every 5 minutes. The data consists of netflow records $D = (F, P, T)$, which indicate that at time $T$, there is a flow with an IP header 5-tuple $F$, and size of $P$ packets. We use $m$ 4-universal hash functions $h_1, h_2,...h_m$ to construct sketches of size $s$. The 4-universal hash function is a special case of $k$-universal hash function. For a $k$-universal hash function, the probability that two different keys both hash to the same value for any $k$ hash functions is exponentially small in $k$. The first 21 bits of source IP address concatenated with the first 21 bits of destination IP address is used as hash key. The $Defeat$ algorithm consists of the following steps.

\begin{itemize}
\item  Compute local sketches: Each router $R_i$ collects the netflow records of the traffic arriving to the network, and constructs $m$ sketches for each of the four features: source and destination IP addresses (SIP and DIP), source and destination port numbers (SP and DP). Entropy is used to measure anomalous distributions. For each of the SIP, DIP, SP and DP, $m$ histogram-sketches are constructed. For each flow $j$ with record $D = (F, P, T)$, we have $k = h_j(HashKey(F))$. A record $(SIP, P)$ is added to the SIP histogram-sketch $j$ at the $k$-th entry. Similar procedures are performed for DIP, SP and DP histogram-sketches. We denote these four histogram-sketches as $H_{i,f,j}, i = 1, 2,...n, f \in (SIP, DIP, SP, DP), j = 1, 2,...m$.
\item  Compute global sketches: In this step, the local histogram-sketches $H_{i,f,j}$ are summed to form global sketches $H_{f,j}$. The histogram associated with $H_{f,j}$ can be treated as an empirical distribution, whose entropy we can compute using (\ref{eqn:pca_entropy}). The result is a $4\times m$ matrix $X_{f,j}$, which constitutes the input to the subspace method. There is one matrix for each combination of feature and hash function.
\item  Detect anomalies: the multi-way subspace method described in the previous subsection is applied to each $X_{f,j}$. The outcome is a bit vector $b$, whose $j$-th bit is 1 if an anomaly is detected using hash function $h_j$. Using a voting approach, we declare an anomaly if $l$ bits out of the total $m$ bits in $b$ is 1, which makes the detection robust to false positives.

\item  Identify anomalies: In this step, we want to identify the IP flows associated with the anomaly. First, we identify the sketch entries in $X_{f,j}$ that are anomalous using the greedy identification heuristic introduced in \cite{lakhina2005}. Then, we determine the set of keys that were mapped to the anomalous sketches. The intersection of these keys identifies the anomalous IP flows.
\end{itemize}
Evaluations using traffic data from two tier-1 backbone networks show that $Defeat$ provides strong detection and identification performances \cite{li2006}.

\subsection{Anomaly Detection for Correlated Flows}
In \cite{silveira2010}, a anomaly detection scheme called ASTUTE is proposed to detect correlated anomalous flows. ASTUTE has low computational complexity and is motivated by the following facts. When there are many flows multiplexed on a non-saturated link, the flows' volume changes over short periods of time have a tendency to cancel each other out so that the average change across flows is close to zero. This phenomena is called a short-timescale uncorrelated-traffic equilibrium (ASTUTE). ASTUTE holds where the flows are nearly independent to each other. ASTUTE does not hold where there are several, potentially small, correlated flows. These flows increase or decrease their volumes at the same time, even when they do not share common 5-tuple features such as IP addresses, port numbers, and protocols. Such behavior is present in many types of traffic anomalies, such as port scanning, DDoS attacks, link outages, routing changes, etc. Compared with other anomaly detection schemes, ASTUTE has the following three advantages:
\begin{itemize}
\item  It doe not require a training phase, which implies low computational complexity and immunity to data-poisoning.
\item  It specializes in detecting a class of anomalies, i.e., strongly correlated anomalous flows, where it performs better than other detection schemes.
\item  It provides information about the anomaly that can be used in anomaly classification.
\end{itemize}
In the following, we first describe the traffic model, and then describe anomaly detection.

\subsubsection{The Traffic Model}
We assume time is divided into fix-length slots. Let $x_{f,i}$ denote the number of bytes in flow $f$ during the $i$th time slot. We assume that the flows traversing a link are generated by a discrete-time market point process \cite{daley1988}, which determines both the flow duration and the flow volume per time slot. Each flow $f$ is determined by three parameters.
\begin{itemize}
\item  $s_f$: the slot when the flow starts.
\item  $d_f$: the number of slots the flow lasts.
\item  $X_f =(x_{f, s_f},...x_{f, s_f + d_f - 1})$: a vector of volumes for all the slots of the flow.
\end{itemize}
In ASTUTE, two assumptions are made.
\begin{itemize}
\item  (A1) Flow independance: A flow's characteristics $s_f, d_f, X_f$ are independent of those of other flows.
\item  (A2) Stationarity: The distributions of the flow arrival process and the marked point process do not change over time.
\end{itemize}
There are two scenarios where flow independence does not hold: 1) Multiple flows in the same session are correlated. 2) Flows in a saturated link are correlated, since they share the same queue. It has been shown that aside the two scenarios just mentioned, the dependencies across real traffic flows are typically very weak\cite{barakat2002, hohn2003}. One explanation is that most backbone links are non-saturated as they are over-provisioned by design. Stationarity depends on the size of the time slot. Traffic typically exhibits non-stationarity at long timescales, e.g., daily, weekly, etc, whereas it exhibits stationarity at short timescales, e.g., less than an hour \cite{cao2001, nychis2008}

We consider two consecutive time slots $i$ and $i + 1$. Let $F$ denote the set of flows active in slot $i$ or $i + 1$. Let $\delta_{f,i} = x_{f,i+1} - x_{f,i}$ denote the volume change from $i$ to $i + 1$, and let $\Delta_{i}$ denote the set of $\delta_{f,i}$'s for all $f$. We have following result.
\begin{itemize}
\item  (R1) If both (A1) and (A2) hold, the variables in $\Delta_i$ are zero mean, i.i.d. random variables.
\end{itemize}
The above result forms the basis for ASTUTE anomaly detection.
\subsubsection{Anomaly Detection}
To detect anomalies, we use traffic on non-saturated links in short-timescale slots. Let $\hat{\delta}_i$ and $\hat{\sigma}_i$ denote the sample mean and standard deviation, respectively, i.e.,
        \begin{equation}
		\label{eqn:pca_ast_mstd}
			\hat{\delta}_i = \sum_{f=1}^{|F|}\frac{\delta_{f,i}}{|F|}\quad \hat{\sigma}_i = \sqrt{\sum_{f=1}^{|F|}\frac{(\delta_{f,i} - \hat{\delta}_i )^2}{(|F| - 1)}}.
		\end{equation}
For large enough $|F|$, $\hat{\delta}_i$ has a $(1 - p)$-confidence interval given by
        \begin{equation}
		\label{eqn:pca_ast_mstd}
			I_{\hat{\delta}_i} = [\hat{\delta}_i - \frac{z_p \hat{\sigma}_i}{\sqrt{|F|}}, \hat{\delta}_i + \frac{z_p \hat{\sigma}_i}{\sqrt{|F|}}]
		\end{equation}
where $z_p$ is the $(1 - p/2)$-quantile of the Gaussian distribution. We declare an anomaly if the confidence interval does not contain zero.

In \cite{silveira2010}, ASTUTE is compared with two well know anomaly detection schemes: Kalman filter \cite{soule2005} and wavelet \cite{barford2002}. ASTUTE is more effective than Kalman filter and wavelet in detecting anomalies that have a large number of flows, especially when the aggregate volume of these flows is small. ASTUTE can detect anomalies that have one or two magnitudes fewer packets than those detected by Kalman filter and wavelet. However, ASTUTE performs worse than Kalman filter and wavelet in detecting anomalies that involve a few large flows.

\subsection{A Communication-Efficient Approximation Scheme}
A communication-efficient PCA-based approximation scheme was proposed in \cite{huang2007}. The scheme avoids the expensive centralization of data processing by performing intelligent filtering at the distributed monitors. The filtering reduces the communications cost but can cause detection errors. The scheme selects the filtering parameters at local monitors such that the errors are bounded at the user-specified level. Thus, the network operator can explicitly balance the tradeoff between the communications cost and the detection accuracy. In the following, we first describe the approximation scheme, and then describe the parameter selections.
\subsubsection{The Approximation Scheme}
In this scheme, there is a set of $n$ distributed monitors $M_1,M_2,...M_n$, each of which collects locally-observed time-series data. There is a central coordinator node that collects data from the distributed monitors and makes detection decisions about network traffic volume anomaly. Each monitor $M_i$ collects a new data point $x_i(t)$ at time step $t$, and sends the data to the coordinator. The coordinator maintains the data within a time window of size $m$ for each monitor's data, and then organizes the data into an $m\times n$-dimensional matrix $X$. The coordinator makes detection decisions based on $X$.

The monitors send descriptions of their time-series data, and only send more measurements or summaries when the triggering condition is met, which is given by (\ref{eqn:pca_traf_cret}). The scheme consists of two parts: 1) the monitors process their data and apply filtering to avoid unnecessary updates to the coordinator; and 2) the coordinator makes global decisions and gives feedback to the monitors based on the updates.

Let $y_i(t)$ denote the approximate representation at the coordinator of monitor $i$'s data $x_i(t)$. We can consider $y_i(t)$ as a predication of $x_i(t)$, which can be the latest $x_i(t)$ sent by monitor $i$, or the value derived by some more sophisticated estimation models \cite{cormode2005, jain2004}. In the following, we describe the protocols at the monitors and the coordinators.

\emph{The monitor protocol}: The monitor $M_i$ continuously tracks the deviation of $x_i(t)$ from its prediction $y_i(t)$, which is given by
        \begin{equation}
		\label{eqn:pca_approx_e}
			e_i(t) = x_i(t) - y_i(t)
		\end{equation}
and checks the condition
        \begin{equation}
		\label{eqn:pca_approx_condition}
			|e_i(t)| \leq \delta_i.
		\end{equation}
If the condition does not hold, the monitor sends an update to the coordinator, which includes $x_i(t)$ and an up-to-date prediction $y_i(t)$, and sets $e_i(t)$ to zero.

\emph{The coordinator protocol}: Given user-specified false alarm probability deviation $\mu$, the coordinator has two tasks: 1) performing anomaly detection based on $y_i(t)$; and 2) computing the filtering parameter $\delta_i$ for each monitor. The coordinator keeps a perturbed version $\hat{X}$ of the global data matrix $X$. The PCA at the coordinator is performed on the the perturbed version of the covariance matrix as given below
        \begin{equation}
		\label{eqn:pca_approx_covariance}
			\hat{C} = \frac{1}{m} \hat{X}^T\hat{X} = C + \Delta.
		\end{equation}
The magnitude of the perturbation matrix $\Delta$ is determined by the filtering parameter $\delta_i$. We can bound $\Delta$ through the control of $\delta_i$.

The coordinator protocol is as follows. Each time the coordinator receives updates from one or more monitors, it carries out the following:
\begin{itemize}
\item  Creates a new row of data $\hat{x} = [\hat{x}_1, \hat{x}_2,...\hat{x}_n]$, where $\hat{x}_i$ is either the update from monitor $i$ or the corresponding prediction $y_i(t)$.
\item  Updates its global view $\hat{X}$ by replacing the oldest row with $\hat{x}$.
\item Using $\hat{X}$, re-computes PCA, the projection matrix $\hat{P}$, and the threshold $\hat{\sigma}_{\alpha}^2$.
\item Performs anomaly detection using $\hat{x}, \hat{P}$, and $\hat{\sigma}_{\alpha}^2$; and triggers an alarm if the following holds
        \begin{equation}
		\label{eqn:pca_approx_alarm}
			\|(1 - \hat{P})\hat{x}\|^2 > \hat{\sigma}_{\alpha}^2.
		\end{equation}
\end{itemize}
The algorithm is listed in Algorithm \ref{alg:monitor}.
		\begin{algorithm}
        	   \begin{algorithmic}[1]
                \Statex Monitor Algorithm
				\Statex Input: Monitor index $i$, local filtering parameter $\delta_i$.  	
				\While{true}
				\State $t$ := current time
				\State $e_i(t) = x_i(t) - y_i(t)$
				\State If ($|e_i(t)| > \delta_i$) then send update ($i, x_i(t), y_i(t)$) to the coordinator, and set $e_i(t) = 0$
				\State If (new filter parameter $\delta_i'$ is received from the coordinator) then set $\delta_i = \delta_i'$
                \EndWhile
                \end{algorithmic}
                \caption{Algorithms for the Monitors and the Coordinator}
                \label{alg:monitor}	
			\begin{algorithmic}[1]
                \Statex Coordinator Algorithm
				\Statex Input: false alarm probability deviation $\mu$.  	
                \While{true}
				\State Make a new row of data $\hat{x} = [\hat{x}_1, \hat{x}_2,...\hat{x}_n]$
				\State Replace the oldest row of $\hat{X}$ using $\hat{x}$
                \For{each (monitor update ($i, x_i(t), y_i'(t)$))}
			    \State Set local prediction $y_i(t) = y_i'(t)$
				\State Set $\hat{x}_i(t) = x_i(t)$
				\EndFor				
                \State Using $\hat{X}$, re-compute PCA, $\hat{P}$, and $\hat{\sigma}_{\alpha}^2$
	            \State if ($\|(1 - \hat{P})\hat{x}\|^2 > \hat{\sigma}_{\alpha}^2$) then trigger anomaly
                \State Compute new optimal filtering parameters $\{\delta_i\}$ based on $\mu$, and disseminate $\{\delta_i\}$

                \EndWhile		
			\end{algorithmic}
		\end{algorithm}
\subsubsection{Parameter Selections}
Let $\alpha$ denote the false alarm probability of using the exact scheme of (\ref{eqn:pca_traf_cret}). Let $\hat{\alpha}$ denote the false alarm probability of the approximation scheme. The false alarm probability deviation $\mu$ specifies the tolerance, to which $\alpha$ and $\hat{\alpha}$ are allowed to differ, i.e.,
        \begin{equation}
		\label{eqn:pca_tolerance}
			\hat{\alpha} - \alpha < \mu.
		\end{equation}

Let $\lambda_i$ and $\hat{\lambda}_i, i = 1,2,...n$ denote the eigenvalues of the covariance matrix $C = \frac{1}{m} X^TX$ and its perturbed version $\hat{C} = \frac{1}{m} \hat{X}^T\hat{X}$. For the metric of errors between two sets of eigenvalues, we use the $l_2$ aggregate eigen-error $\epsilon^*$ defined as
        \begin{equation}
		\label{eqn:pca_epsilon}
			\epsilon^* = \sqrt{\frac{1}{n}\sum_{i=1}^n(\hat{\lambda}_i - \lambda_i)^2}.
		\end{equation}

We proceed in two steps: 1) given $\mu$, determine an upper bound on $\epsilon^*$; 2) given $\epsilon^*$, determine filtering parameter $\delta_i$.

\emph{Step 1: From false alarm deviation $\mu$ to eigen-error $\epsilon^*$}: There is no closed-form solution for eigen-error $\epsilon^*$. We use binary search to otain $\epsilon^*$, starting from an initial guess and then computing our estimate for the resulting $\mu^*$. The algorithm is listed in Algorithm \ref{alg:eigen-error}.
        \begin{algorithm}
        	\begin{algorithmic}[1]
                \Statex Input: Deviation $\mu$; desired approximation factor (err) for eigen-error.  	
				\Statex $\epsilon_l = 0; \epsilon_u = \bar{\lambda}$  //search range for $\epsilon$
                \While{true}
				\State $\epsilon = 0.5(\epsilon_l + \epsilon_u)$
				\State $\eta_Z = \text{MonteCarloSampling}(\epsilon)$
                \State $\mu^* = \text{Pr}[c_\alpha - \eta_Z < N(0,1) < c_\alpha]$, where $c_\alpha$ is the $(1 - \alpha)$-percentile of the standard normal distribution and $N(0, 1)$ is a standard normal random variable
				\State if ($\mu_* > \mu$) then $\epsilon_u = \epsilon \quad\text{else}\quad \epsilon_l = \epsilon$
				\EndWhile				
                \State return($\epsilon$)
			\end{algorithmic}
            \caption{Algorithms for Searching for Eigen-error}
            \label{alg:eigen-error}
		\end{algorithm}

Next, we describe how to estimate $\mu$. We will use the following random vector
        \begin{equation}
		\label{eqn:pca_Y}
			Y = \frac{\phi_1(\|(1-P)x\|^2/\phi_1)^h_0 - \phi_2h_0(h_0-1)/\phi_1^2}{\sqrt{2\phi_2h_0^2}}
		\end{equation}
where $h_o, \phi_i$ are given by (\ref{eqn:pca_q_test}). The random vector $Y$ normalizes $\|(1 - P)\|^2$ and follows normal distribution \cite{jensen2004}. To perform detection on $\|(1 - P)x\|$ with false alarm probability of $\alpha$, the threshold $\sigma_{\alpha}^2$ can be obtained by a high-order complex function of $\lambda_{k+1},\lambda_{k+2},...\lambda_{n}$ \cite{jackson1979}. Based on (\ref{eqn:pca_Y}), we obtain the false alarm probability of the original PCA-based detector as
        \begin{equation}
		\label{eqn:pca_fap1}
			\text{Pr}[|(1 - P)\|^2 > \sigma_{\alpha}^2] = \text{Pr}[Y > c_\alpha] = \alpha
		\end{equation}
where $c_\alpha$ is the $(1 - \alpha)$-percentile of the standard normal distribution.

Let $\hat{Z} = \|(1 - \hat{P})\hat{x}\|$, and $\eta_Z$ denote an upper bound on $\|\hat{Z} - Z\|$. The deviation of the false alarm probability can be expressed by
        \begin{equation}
		\label{eqn:pca_fap2}
			\mu = \text{Pr}[c_\alpha - \eta_Z < N(0, 1) < c_\alpha]
		\end{equation}
where $N(0, 1)$ is a standard normal random variable. To estimate $\eta_Z$, we use a Monte Carlo sampling technique, details of which can be found in \cite{huang2006}.

\emph{Step 2: From eigen-error $\epsilon^*$ to filtering parameter $\delta_i$}: Let $W$ denote the filtering error matrix, i.e., $W = X - \hat{X}$. Because of the filtering methods used, all the elements of the column vectors $W_i$ are bounded within the interval $[-\delta_i, \delta_i]$. The following assumptions are made, which are standard in the Stochastic Matrix Theory.
\begin{itemize}
\item  The column vectors $W_1, W_2,...,W_n$ are independent and radially symmetric $m$-dimensional random vectors. In other words, their projection on a sphere is uniformly distributed.
\item For each $i = 1, 2,...,n$, all elements of $W_i$ are i.i.d. random variables with zero mean and variance of $\sigma_i^2$.
\end{itemize}

Let $\bar{\lambda} = \frac{1}{n}\sum\hat{\lambda}_i$ denote the average of the perturbed eigenvalues of $\hat{C}$. It can be shown that $\epsilon^* \leq \epsilon$ with probability larger than $1 - o(m^{-3})$ \cite{huang2006}, with $epsilon$ given below
        \begin{equation}
		\label{eqn:pca_epsilon}
			\epsilon = 2\sqrt{\frac{\bar{\lambda}}{m}\sum_{i=1}^n\sigma_i^2} + \sqrt{(\frac{1}{m} + \frac{1}{n})\sum_{i=1}^n\sigma_i^4}.
		\end{equation}
Similar results also hold for the eigen-subspace $1 - P$ and individual eigenvalues. Given a tolerable eigen-error $\epsilon$, we can use (\ref{eqn:pca_epsilon}) to solve for filtering parameter $\delta_i$. Different techniques can be employed to quantify the relationship between $\delta_i$ and $\sigma_i$, which are listed below.
\begin{itemize}
\item  Homogeneous filtering parameter allocation: the uniform distribution method: This is a simple method that often works well in practice. We assume the filtering parameters are i.i.d. random variables, which implies $\sigma_i = \delta_i^2/3$. The filter parameters are homogeneous, i.e., $\sigma_i = \sigma$. We can solve (\ref{eqn:pca_epsilon}) directly and obtain
        \begin{equation}
		\label{eqn:pca_sigma}
			\sigma = \frac{\sqrt{3\bar{\lambda}n + 3\epsilon\sqrt{m^2 + mn}} - \sqrt{3\bar{\lambda}n}}{\sqrt{m + n}}.
		\end{equation}
\item  Homogeneous filtering parameter allocation: the local variance estimation method: The assumption of uniform distribution may be violated in some scenarios. In such case, we can estimate local error variances $\delta_i, \sigma$ directly from the data. We can perform the estimation by fitting a (e.g., quadratic) function of $\delta$ using a recent window of observations. These functions are sent to the coordinator and plugged into (\ref{eqn:pca_epsilon}) to solve for a new $\delta$.
\item  Heterogeneous filtering parameter allocation: In this method, local filtering parameters $\delta_1, \delta_2,....,\delta_n$ can differ from one another, and can adapt dynamically to local stream characteristics. The message update frequency of monitor $M_i$ is a function $f_i(\delta_i)$. The heterogeneous filtering parameter allocation can be formulated as an optimization problem as follows
        \begin{equation}
		\label{eqn:pca_heterogeous_allocation}
			\text{Minmize}\quad\sum_{i=1}^nf_i(\delta_i)\quad \text{such that}\quad 2\sqrt{\frac{\bar{\lambda}}{m}\sum_{i=1}^n\sigma_i^2/3} = \epsilon
		\end{equation}
    where the second summand in (\ref{eqn:pca_epsilon}) is omitted, since it is typically an order of magnitude smaller than the first one.
\end{itemize}

It was shown in \cite{huang2007} that using the above methods anomaly detection can be performed with 80-90\% reduction in communications cost. Moreover, the system scales gracefully as the number of monitors is increased, and the coordinator's input data rate an order of magnitude more slowly than a system that sends all monitoring data.
\subsection{Problems and Solutions with Applying PCA for Anomaly Detection}
It has been shown that PCA is sensitive to parameter settings \cite{ringberg2007}. Reference \cite{brauckhoff2009b} shows that the problem with PCA is that it does not consider temporal correlation of the data, and it provides a solution to the problem, the details of which is given below.
\subsubsection{Problems with the Application of PCA for Anomaly Detection}
For the application of PCA to be valid, two conditions must hold for the data: 1) they must be linear, i.e., they can be represented by a linear combination of independent random variables. 2) The mean and covariance provide sufficient statistics for the data, i.e., the mean and covariance entirely determine the joint probability distribution. When the two conditions are met, the most suitable basis is the one that maximizes the variance of each projected component, and PCA is most effective. One instance where the two conditions hold is that the random variables are jointly Gaussian. There are quite few examples in the literature where PCA was applied when the two conditions were not met.

We provide a brief review of the theory of PCA here to prepare for the discussion in the next section. Let $x(t) = \{x_1(t), x_2(t),...x_N(t)\}$ denote an $N$-dimensional vector of zero-mean stationary stochastic processes. The random vector $x$ can be decomposed onto the principal axes as below
        \begin{equation}
		\label{eqn:pca_decomp2}
			x = \sum_{i=1}^N u_i v_i
		\end{equation}
where $v = (v_1, v_2,...v_N)$ are the ortho-normal principal axes of $x$, and $u = (u_1, u_2,...u_N)$ are principal components of $x$, which are uncorrelated with each other. The principal axes $v_i$'s are the eigen-vectors of the covariance matrix $C = \text{E}[xx^T]$ of $x$, i.e,
        \begin{equation}
		\label{eqn:pca_corr_eigen2}
			Cv_i = \lambda_i v_i.
		\end{equation}
\subsubsection{Extension of PCA}
Let $\sigma_{i,j}(\tau) = \text{E}[x_i(t)x_j(t-\tau)]$, defined on the interval $[a, b]$. The multi-dimension Karhunen-Loeve theorem \cite{gray2005} provides an KL expansion as described below
        \begin{equation}
		\label{eqn:pca_kl_exp}
			x_l(t) = \sum_{i=1}^N \sum_{j=1}^\infty u_{i,j}^l v_{i,j}(t)
		\end{equation}
where $u_{i,j}^l$'s are the pairwise independent random variables and $v_{i,j}(t)$'s are the pairwise orthogonal deterministic basis. The above equation is the equivalent to (\ref{eqn:pca_decomp2}). The equivalent of (\ref{eqn:pca_corr_eigen2}) is given by
        \begin{equation}
		\label{eqn:pca_corr_eigen3}
			\sum_{i=1}^N \int_a^b \sigma_{i,k}v_{i,j}(s-t) ds = \lambda_{k,j} v_{k,j}(t).
		\end{equation}

The KL expansion provides an extension to the PCA. It considers both the temporal correlation between $t$ and $t + \tau$ and the spatial correlation between $x_i$ and $x_j$. Not considering the temporal correlation would cause errors as described by \cite{ringberg2007}.

We take samples at discrete times $kT$ for a total of $K$ time intervals. We assume that the covariance $\sigma_{i,j}(\tau)$ is negligible for $\tau > JT$. The Galerkin method is used to generate a set of  $N\times J$-dimensional eigen-vectors $v_{i,j}(k) = v_{i,j}(kT)$ \cite{kunisch2002}. We obtain a discrete version of the KL expansion as below
        \begin{equation}
		\label{eqn:pca_kl_exp2}
			x_l(k) = \sum_{i=1}^N \sum_{j=1}^J u_{i,j}^l v_{i,j}(k).
		\end{equation}

We neglect the smaller terms of the KL expansion and obtain an approximation of $X_l[k]$ as follows
        \begin{equation}
		\label{eqn:pca_kl_exp3}
			\hat{x}_l(k) = \sum_{i=1}^L \sum_{j=1}^M u_{i,j}^l v_{i,j}(k)
		\end{equation}
where $L < N$ and $M < J$. The above approximation has the smallest error variance Var[$X(t)-\hat{X(t)}$] among all approximations defined over an $L\times M$-dimensional linear space. This provides the theoretical basis for using KL expansion as a non-parametric and generic technique to model a class of processes, where there is no guarantee of linearity and sufficiency of mean and variance.

The above method was evaluated using data from a medium-sized ISP \cite{brauckhoff2009b}. It was found the anomaly detection results are much improved by considering the temporal correlations.


    \section{Sketch-Based Approaches}
		\label{section:sketch}
		In this section, we first introduce the basic anomaly detection scheme using sketches in subsections A, B, and C. In subsection D, the sketch-based scheme is extended to detect high-volume traffic clusters. In subsection E, the sketch method is combined with non-Gaussian multi-resolution statistical detection produces, with the former reducing data dimensionality and the latter detecting anomaly at different aggregation levels. 

Next, we provide an overview of the sketch method. Data stream computation is an effective technique to process massive data streams online. Using this technique, data is processed exactly once. A good survey paper can be found in \cite{muthukrishnan2003}. One particular technique of data stream computation is the sketch, a probabilistic summary method that uses random projections.

We use the Turnstile model \cite{muthukrishnan2003} to describe the data stream method. Let $I = I_1,I_2,...$ denote an input stream that arrives sequentially. Each item $I_i = (a_i, u_i)$ is composed of a key $a_i \in \{0, 1,..., u - 1\}$, and an update $u_i \in R$. Associated with each key $a_i$ is a signal $A[a_i]$. The arrival of item $I_i$ causes an update as follows
        \begin{equation}
		\label{eqn:sketch_update1}
			A[a_i] = A[a_i] + u_i.
		\end{equation}
In the context of network anomaly detection, we can use one or more fields in the IP header as the key. The update can be the packet size, the total number of bytes or packets in a traffic flow. In the following, we use the destination IP address as the key.

A sketch-based anomaly detection scheme was proposed in \cite{krishnamurthy2003}, which consists of three modules: sketch, forecasting and change detection. The sketch module generates a sketch to summarize all the updates. The forecasting module generates a forecast sketch based on the observed sketches in the past, and also a forecast error sketch. The change detection module uses the error sketch to detect changes. The sketch-based approach was extended to online identification of hierarchical heavy hitters in \cite{zhang2004}. Details are given below.
\subsection{The Sketch Module}
Given input stream $(a_i, u_i)$, we compute a sketch corresponding to key $a$ given by
        \begin{equation}
		\label{eqn:sketch_v_a}
			v_a = \sum_{i\in A_a} u_i, \quad A_a = \{i |a_i=a\}.
		\end{equation}
Also, we define the second moment as
        \begin{equation}
		\label{eqn:sketch_f_2}
			F_2 = \sum_a v_a^2.
		\end{equation}

We use a particular variant of sketch called k-ary sketch, which is a $K\times H$ matrix $S$, whose elements are sketches. We use $H$ hash functions. Column $i$ of $S$ is associated with a 4-universal hash function $h_i$ \cite{carter1979, wegman1981}. As mentioned in the previous section, the 4-universal hash function is a special case of $k$-universal hash function. For a k-universal hash function, the probability that two different keys both hash to the same value for any $k$ hash functions is exponentially small in $k$. Different hash function uses different seeds for random number generators. There are four basic operations for the k-ary sketches as described below:
\begin{itemize}
\item  Update: Similar to (\ref{eqn:sketch_update1}), we update the sketch $S$ once an item $u$ arrives as follows
        \begin{equation}
		\label{eqn:sketch_update2}
			S_{i,h_i(a)} = S_{i,h_i(a)} + u, \forall i.
		\end{equation}
\item  Estimate $v_a$: We estimate $v_a$ as follows
        \begin{equation}
		\label{eqn:sketch_estimate1}
			v_a = \underset{i}{\text{median}}(v_{a,h_i})
		\end{equation}
where
        \begin{equation}
		\label{eqn:sketch_estimate2}
			v_{a,h_i} = \frac{S_{i,h_i(a)} - sum_s/K}{1 - 1/K}
		\end{equation}
and
        \begin{equation}
		\label{eqn:sketch_estimate3}
			sum_s = \sum_{j=1}^K S_{0,j}.
		\end{equation}
The median() function returns the median among the inputs. In other words, the hash function $h_i()$ maps the destination IP address to a value $h_i(a)$. Then the $h_i(a)$-th column of matrix $S$ is updated.
It can be shown that each $v_{a,h_i}$ is an unbiased estimator of $v_a$ with variance inversely proportional to $K - 1$.
\item  Estimate $F_2$: We estimate $F_2$ as follows
        \begin{equation}
		\label{eqn:sketch_f2_estimate1}
			F_2 = \underset{i}{\text{median}}(F_{2, h_i})
		\end{equation}
where
        \begin{equation}
		\label{eqn:sketch_f2_estimate2}
			F_{2,h_i} = \frac{K}{K - 1}\sum_{j=1}^K S_{i,j}^2 -  \frac{sum_s^2}{K - 1}.
		\end{equation}
Again, it can be shown that each $F_{2,h_i}$ is an unbiased estimator of $F_2$ with variance inversely proportional to $K - 1$.
\end{itemize}
\subsection{The Forecasting Module}
The forecasting module generates a forecast sketch $S_f(t)$ based on the observed sketches $S_o$ in the past, which is the sketches obtained in the previous subsection. We use six models for the univariate time series forecasting as described below:
\begin{itemize}
\item  Moving Average (MA): In this model, equal weights are assigned to all past samples. It has an integer parameter $W \geq 1$ that determines the number of past time intervals used in the forecasting. The forecast sketch is computed as follows
        \begin{equation}
		\label{eqn:sketch_ma}
			S_f(t) = \frac{\sum_{i=1}^WS_o(t - i)}{W}.
		\end{equation}
\item S-shaped Moving Average (SMA): In this model, more weights are given to more recent samples, and the forecast sketch is computed as follows
        \begin{equation}
		\label{eqn:sketch_sma}
			S_f(t) = \frac{\sum_{i=1}^Ww_iS_o(t - i)}{\sum_{i=1}^Ww_i}.
		\end{equation}
In the implementation, equal weights are given for the most recent half of the forecasting window $W$, and linearly decaying weights for the earlier half. The reader is referred to \cite{floyd2000} for the choices of the weights.
\item  Exponentially Weighted Moving Average (EWMA): In this model, the forecast for time $t$ is the weighted average of the previous forecast and the current observed sample $S_o(t-1)$, i.e.,
        \begin{equation}
        \label{ean:sketch_emwa}
        S_f(t) =
        \begin{cases}
        \alpha S_o(t - 1) + (1 - \alpha)S_f(t - 1)  &  t > 2\\
        S_o(1) & t = 2
        \end{cases}
        \end{equation}
    where $\alpha$ determines how many weights are given to current and past samples.
\item  Non-Seasonal Holt Winters (NSHW): In this model\cite{brockwell1996}, $S_f(t)$ is composed of two components: the smoothing component $S_s(t)$ and the trending component $S_t(t)$ as follows
        \begin{equation}
		\label{eqn:sketch_nshw1}
			S_f(t) = S_s(t) + S_t(t),
		\end{equation}
        \begin{equation}
        \label{ean:sketch_nshw2}
        S_s(t) =
        \begin{cases}
        \alpha S_o(t - 1) + (1 - \alpha)S_f(t - 1)  &  t > 2\\
        S_o(1) & t = 2
        \end{cases}
        \end{equation}
        \begin{equation}
        \label{ean:sketch_nshw3}
        S_t(t) =
        \begin{cases}
        \beta (S_s(t) - S_s(t - 1))+ (1 - \beta)S_t(t - 1)  &  t > 2\\
        S_o(2) - S_o(1) & t = 2.
        \end{cases}
        \end{equation}
\item  AutoRegressive Integrated Moving Average (ARIMA): The model is also called Box-Jenkins model \cite{box1994, box1976}. There are three parameters in this model: the autoregressive parameter ($p$), the number of differencing passes ($d$), and the moving average parameter ($q$). The model can be described by
        \begin{equation}
		\label{eqn:sketch_arima}
			Z_t - \sum_{i=1}^q {MA}_iZ_{t-i} = e_t - \sum_{j=1}^p AR_je_{t-i}
		\end{equation}
    where $Z_t$ is obtained by differencing the original time series $d$ times, $e_t$ is the forecast error at time $t$, $MA_i$ and $AR_j$ are Moving Average and AutoRegression coefficients.

    Typical selection of parameters is: $p \leq 2, d = 0 $ or $1, q \leq 2$. Here, we consider two types of ARIMA models: ARIMA0 of the order ($p \leq 2, d = 0, q \leq 2$) and ARIMA1 of the order ($p \leq 2, d = 1, q \leq 2$). The choice of $MA_i$ and $AR_j$ must guarantee that the resulting models are invertible and stationary, for which a necessary but not sufficient condition is $MA_i$ and $AR_j \in [-2, 2]$ when $p, q \leq 2$.
\end{itemize}
\subsection{The Change Detection Module}
The forecast error sketch $S_e(t)$ is the difference between the observed sketch $S_o(t)$ and the forecasted sketch $S_f(t)$, i.e.,
        \begin{equation}
		\label{eqn:sketch_S_e}
			S_e(t) = S_o(t) - S_f(t).
		\end{equation}
Based on the estimated second moment of $S_e(t)$, an alarm threshold $R_A$ is given by
        \begin{equation}
		\label{eqn:sketch_S_e_threshold}
			R_A = R(\hat{F}_2(S_e(t)))^{1/2}
		\end{equation}
where $R$ is a parameter to be determined by the application. An alarm is raised if the estimated error $S_e(t)$ is above the alarm threshold $R_A$.

It was shown in \cite{krishnamurthy2003} that the sketch-based anomaly detection scheme can detect significant changes in massive data streams. The scheme uses a constant, small amount of memory, and has constant per-record update and reconstruction cost, both of which are very desirable in streaming applications.
\subsection{Online Identification of Hierarchical Heavy Hitters}
The sketch-based approach was extended to online identification of hierarchical heavy hitters in \cite{zhang2004}. Heavy hitters are high-volume traffic clusters. These traffic clusters are often hierarchical in that they occur at different aggregation levels, such as the ranges of IP addresses, and they are also possibly multidimensional, i.e., they may involve a combination of IP header fields, such as IP addresses, port numbers, and protocols. The focus in \cite{zhang2004} is on 1-dimensional and 2-dimensional heavy hitters, i.e., those associated with IP source and/or destination addresses, arguably the most important scenarios.

A precise definition of a hierarchical heavy hitter in terms of sketches is as follows. Let $I = (a_i, u_i)$ denote the traffic input stream, whose keys $a_i$ are drawn from a multidimensional hierarchical domain $D = D_1\times D_2\times...\times D_k$, with $D_i$ having a height of $h_i$. The keys we use here are IP addresses, and the values are traffic volumes in bytes. For any prefix $p = (p_1, p_2,...,p_k) \in D$ of the domain hierarchy, let $elem(D,p)$ denote the set of elements in $D$ that are descendants of $p$. Let $V(D, p) = \sum_{k \in elem(D,p)}u_k$ denote the total values associated with any given prefix, $S$ denote the total sum of values in $I$, i.e., $S = \sum_i u_i$, and $\phi \in [0, 1]$ denote a threshold value. The set of hierarchical heavy hitters is defined as
        \begin{equation}
		\label{eqn:sketch_hhh}
			[p | V(D,p) \geq \phi S].
		\end{equation}
There are two key parameters: $\phi$ and $\epsilon$. To qualify for heavy hitters, the threshold is $\phi S$. On the other hand, $\epsilon S$ is the maximum amount of inaccuracy that is tolerated in the algorithms, which is guaranteed by controlling local decision threshold $T_s$ called split threshold.

In the following, we first describe a baseline heavy hitter detection scheme, followed by 1-dimensional and 2-dimensional detection schemes, ending with a general $n$-dimensional detection scheme.

\subsubsection{Baseline Heavy Hitter Detection}
The baseline scheme is straightforward, but inefficient, and is used as a baseline to evaluate hierarchical detection schemes. Essentially, it transforms the hierarchical heavy hitter detection problem into multiple non-hierarchical detection problems. For a $k$-dimensional hierarchical detection problem with height $h_i$ in the $i$-th dimension, we need to solve $\prod_{i=1}^k(h_i + 1)$ non-hierarchical problems. Two variants of baseline schemes are introduced, which differ in the specific detection algorithms used. Details can be found in \cite{cormode2004, manku2002}.
\begin{itemize}
\item  Sketch based scheme: It uses count-min sketch \cite{cormode2004}. The sketch $S$ is composed of a $K\times H$ matrix, each column of which is associated with a hash function $h_i$. Given a key, the sketch allows us to recover the value with probabilistic bounds on recovery accuracy. It uses a separate sketch for each prefix.
\item  Lossy counting-based scheme: This is a deterministic, single-pass, sampling-based scheme \cite{manku2002}. Let $N$ denote the number of items in the input data stream. This scheme can correctly identify  all heavy hitters whose frequencies exceed $\phi N$.
\end{itemize}
\subsubsection{1-Dimensional Heavy Hitter Detection}
This scheme is trie-based. Trie is often used in IP address lookup \cite{srinivasan1999}. In this scheme, each node of the trie has $2^m$ children. For the ease of exposition, we describe the algorithm using the trie that has $2$ children. Each node $i$ in the trie is associated with a prefix $p^*$, which indicates the path between the root of the trie and the node.

\emph{The trie data structure}: In the data structure of trie listed in Figure \ref{fig:trie1}, array $i.children$ contains the pointers to the children of the node $i$. Field $i.depth$ indicates the depth of node $i$ from the root. Field $i.fringe$ indicates whether node $i$ is a fringe node. Node $i$ is a fringe node if after its creation, we see less than $T_s$ amount of traffic associated with prefix $p$. If not, node $i$ is an internal node. Field $i.volume$ records the traffic volume associated with prefix $p$ after node $i$ is created and before node $i$ becomes an internal node. Field $i.subtotal$ indicates the total traffic volume for the entire subtrie rooted at node $i$, excluding the amount already counted by $i.volume$. Fields $i.miss\_copy$ and $i.miss\_split$ represent estimated traffic volume missed by node $i$, i.e., traffic associated with prefix $p$ but appearing before the creation of node $i$. The $copy\_all$ and $splitting$ rules are used to calculate $i.miss\_copy$ and $i.miss\_split$, respectively, details of which is given below. The last four fields are used to estimate total traffic volume associated with prefix $p$. We will describe the estimation algorithms later.

\begin{figure}[htb!]
		\setlength{\unitlength}{0.14in}
		\centering

		\begin{picture}(12,15)
		
			\put(-5.5,2){\framebox(21,12.3)}
			\put(-4,13) {typedef struct \{ }
			\put(-4,11.7) {trie * child[ ];}
			\put(-4,10.4) {int depth;}
			\put(-4,9.1) {boolean fringe;}
			\put(-4,7.8) {int volume;}
			\put(-4,6.5) {int subtotal;}
			\put(-4,5.2) {int miss\_copy;}
			\put(-4,3.9) {int miss\_split;}
			\put(-4,2.6) {trie; \} }
		\end{picture}
            \caption{The trie data structure}
            \label{fig:trie1}
	
\end{figure}

\emph{Updating the trie}: The trie starts with a single node associated with the zero-length prefix $*$. The $volume$ field associated with this node is incremented with the size of each arriving packet. When the value in this field exceeds $T_s$, the node becomes internal node and a child node is created with the prefix $0*$ or $1*$ that the arriving packet matches. The $volume$ field of the child is updated by the packet size. The trie is updated upon arrival of each new packet. The algorithm is listed in Algorithm  \ref{alg:update_trie}.

\begin{algorithm}
        	\begin{algorithmic}[1]
                \State $i = root $
                \While{true}
				\If {($i.fringe$)}
				    \If {($i.volume + value < T_s$)} $i.volume += value $
                    \State\Return $i.depth - 1$
                    \Else
                    \quad $i.fringe = false$
                        \If {($i.depth = W)} i.subtotal = value$
                        \State\Return $i.depth$
                        \EndIf
                     \EndIf
                \ElsIf{($i.depth = W$)}
                    $i.subtotal += value$
                    \State\Return $i.depth$
                \EndIf
                \State $index = get_Nth\_bit(key, i.depth +1)$
                \State $c = get\_child(i.index)$
                \If {$(c = Null)$}\quad
				        $ c = create\_child(i, index)$
                \EndIf
				\State $n = c$
				\EndWhile				
                \end{algorithmic}
                \caption{Updating the trie}
                 \label{alg:update_trie}
		\end{algorithm}
\begin{figure}[htb!]
 \tiny
		\setlength{\unitlength}{0.21mm}
		\centering

		\begin{tikzpicture}[scale=0.6]
			
			\draw(2.5,3) circle (0.3);
			\draw (2.7,3.2) -- (3.8,4.8) ;
			\draw (5.3,3.2) -- (4.2,4.8) ;
			\draw (5.3,2.8) -- (4.2,1.2) ;
			\draw (5.7,2.8) -- (6.8,1.2) ;
			\draw[dashed] (5.5,3) circle (0.3);
			\draw[dashed] (4,5) circle (0.3);
			\draw(4,1) circle (0.3);
			\draw(7,1) circle (0.3);

			\draw(10.5,3) circle (0.3);
			\draw (10.7,3.2) -- (11.8,4.8) ;
			\draw (13.3,3.2) -- (12.2,4.8) ;
			\draw (13.3,2.8) -- (12.2,1.2) ;
			\draw (13.7,2.8) -- (14.8,1.2) ;
			\draw[dashed] (13.5,3) circle (0.3);
			\draw[dashed] (12,5) circle (0.3);
			\draw(12,1) circle (0.3);
			\draw(15,1) circle (0.3);
			\draw[fill={rgb:black,1;white,2}](10.5,-1) circle (0.3);
			\draw(10.5,-1) circle (0.3);
			\draw (11.8,0.8) -- (10.7,-0.8) ;

			\put(308,110){$0$};
			\put(373,110){$1$};	
			\put(340,140){$5$};
			\put(296,84){$6$};
			\put(382,84){$7$};
			\put(348,53){$0$};
			\put(413,53){$1$};
			\put(339,26){$8$};
			\put(425,26){$9$};
			\put(310,0){$0$};
			\put(297,-31){$5$};

			\put(80,110){$0$};
			\put(145,110){$1$};
			\put(112,140){$5$};
			\put(68,84){$6$};
			\put(154,84){$7$};
			\put(120,53){$0$};
			\put(185,53){$1$};
			\put(111,26){$8$};
			\put(197,26){$9$};

                     \end{tikzpicture}
                     \caption{An example of trie}
                     \label{fig:trie2}
\end{figure}
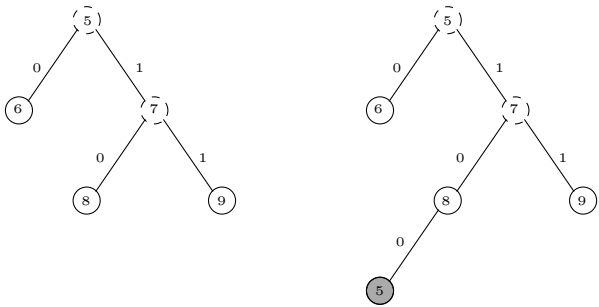

In Figure \ref{fig:trie2}, we use an example to illustrate the algorithm. The arriving packet has a destination IP prefix of $100*$ and a size of 5 bytes. Figure (a) depicts the trie at the time of the packet arrival. The algorithm first performs a longest prefix match and reaches the node associated with the prefix $10*$, which is the node with the value 8. Suppose $T_s = 10$. Adding 5 bytes to the volume field of this node would make its value larger than $T_s$. Thus, a new node is created that is associated with the prefix $100*$ with a value of 5 bytes, as shown in Figure (b).

Since the $volume$ field of any internal node is less than $T_s$, we can ensure that the maximum amount of traffic we miss is at most $\epsilon S$ by setting $T_s = \epsilon S/W$. The time complexity of operations described above is on the same order of magnitude as that of IP lookup, i.e., $O(W)$. For each incoming packet, we update at most one node in the trie, and at most one new node is created assuming the packet size is no more than $T_s$. At each depth, there can be no more than $S/T_s = W / \epsilon$ internal nodes, otherwise the total sum over all the subtries would exceed $S$. So the worst-case memory requirement is $O(W^2/\epsilon)$.

\emph{Reconstructing volumes for internal nodes}: In the building-up of the trie, each incoming packet results in at most one update, which occurs at the node that is most specific to the destination IP prefix of the packet. We reconstruct the volumes of the internal node at the end of the time interval. The reconstruction cost is amortized across the entire time interval.

\emph{Estimating the missed traffic}: Because of using $T_s$ to guide the construction of the trie, the volumes represented in the internal nodes even after the reconstruction are not accurate. To more accurately estimate the volumes, we estimate the missed traffic, of which there are three ways as described below.
\begin{itemize}
\item  Copy\_all: The missed traffic for node $i$ is estimated as the sum of the total traffic seen by the ancestors of node $i$ in the path from the root to node $i$. Copy\_all is conservative in that it copies the traffic to all its descendants, which gives an upper bound for the missed traffic. Since for every internal node, $i.volume \leq T_s$, the estimate given by copy\_all is upper bounded by $depth(node\text{ }i)\times T_s$.
\item  No\_copy: This is the liberal extreme that assumes there is no missed traffic.
\item  Splitting: The missed traffic of node $i$ is split among all its children, where child $j$ receiving an amount proportional of the traffic amount of child $j$. This assumes the traffic patterns are similar before and after the creation of a node. Both copy\_all and splitting can be implemented easily by traversing the trie in a top-down fashion.
\end{itemize}

\emph{Detecting heavy hitters}: After we have an estimate of the missed traffic, we can combine it with the traffic volume and use the sum as input for heavy hitter detection. The accuracy depends on the rule we used. Copy\_all guarantees there is no false negative but there can be some false positives. No\_copy is exactly the opposite. Splitting has fewer false positives than copy\_all and fewer false negatives than no\_copy.
\subsubsection{2-Dimensional Heavy Hitter Detection}
The detection scheme is an adaptation of the cross-producting technique \cite{srinivasan1998}, which was originally used for packet classification. The basic idea is to perform 1-dimensional heavy hitter detection for each of the dimensions, and to use the lengths associated with the longest prefix match nodes in each dimension as indices into a data structure that holds the volumes for the 2-dimensional heavy hitters.

There are three data structures. Two tries are used to maintain two 1-dimension information. A $W\times W$ array $H$ of hash tables is used to keep track of the 2-dimensional tuples. A tuple $(p_1, p_2)$ consists of the longest matching prefix in both dimensions. The array is indexed by the prefix lengths of the prefixes $p_1$ and $p_2$. In the case of IPv4, for a 1-bit trie-based scheme, $W = 32$.

\emph{Updating the data structure}: When a packet arrives, we first update the individual 1-dimensional tries, which returns the longest matching prefix in each dimension $p_1$ and $p_2$ with lengths of $l_1$ and $l_2$ respectively. The two lengths are used as indices to identify the hash table $H_{l_1, l_2}$, in which $(p_1, p_2)$ is used as a lookup key. The volume field associated with the key is incremented. This process is repeated for every incoming packet.

For each incoming packet, three update operations are performed: one operation in each of the two 1-dimensional tries, and one operation in at most one of the hash tables. The memory requirement in the worst case is $O(W^4/\epsilon^2)$, due to cross-producting. In practice, the actual memory requirement is much lower.

\emph{Reconstructing volumes for internal nodes}: We add the volume for each element in the hash tables to all its ancestors, which can be implemented by scanning all the hash elements twice. In the first pass, for every entry $e$ represented by key $(p_1, p_2)$ with lengths $(l_1, l_2)$, we add the volume associated with $e$ to its left parent represented by key $(ancestor(p_1), p_2)$ with lengths $(l_1-1, l_2)$. We start with entries with the largest $l_1$, ending with entries with smallest $l_1$. In the second pass, we add the volume to the right parent represented by key $(p_1, ancestor(P_2))$ with lengths $(l_1, l_2-1)$. This time, we start from entries with largest $l_2$, ending with those with smallest $l_2$.

{\em Estimating the missed traffic}: For each key, we traverse the individual tries to find the prefix represented by the key, and return the missed traffic estimate by applying either copy\_all or splitting rules. The missed traffic is then estimated as the maximum of the two estimates from individual trie. The maximum preserves the conservativeness of copy\_all.

The cross-producting technique is efficient in time, but it can be memory intensive. This downside can be ameliorated by using the techniques of grid-of-tries and rectangle search \cite{srinivasan1998}. A couple of other optimizations can also be made, such as lazy expansion, compression, etc.  The techniques can also be adapted to change detection. Results from evaluations using real Internet traces from a tier-1 ISP indicate these techniques are remarkably accurate and efficient in resource usage. For details, the reader is referred to \cite{zhang2004}.
\subsection{Anomaly Detection Using Sketches and Non-Gaussian Multi-resolution Statistical Detection Procedures}
In \cite{dewaele2007}, an anomaly detection and characterization scheme is proposed, which uses both the sketches and non-Gaussian multi-resolution statistical detection procedures. The former reduces the dimensionality of the data, and the latter detects anomaly at different aggregation levels. The scheme is capable of detecting both short-lived and long-lived low-intensity anomalies, and uses only single-link measurements.

The anomaly detection scheme consists of six steps as described below.

\begin{itemize}
\item  Step 1: Sketches. Sketches are taken for each time-widow of duration $T$. Let $(t_i, x_{i,l}, l = 1, 2, 3, 4)$ denote packet arrival time, source and destination IP addresses (SIP, DIP), and source and destination port numbers (SP, DP) for each packet $i = 1, 2,...I$. Let $h_n, n = 1,2,...N$ denote independent $k$-universal hash functions. Let $M$ stands for the identical size of hash tables, $K_i$ the hashing key. In our case $K_i = SIP_i$ or $K_i = DIP_i$. The original trace $(t_i, x_{i,l})$ is spit onto $M$ sub-traces $(t_i, m = h_n(K_i))_{n,m}$, each corresponding to a particular $h_n$.
\item  Step 2: Multi-resolution aggregation. The $M$ sub-traces $(t_i, m = h_n(K_i))_{n,m}$ are aggregated jointly over a range of levels $\delta_j, j = 1,2,...J$ to form the $X_{\delta_j}^{n,m}$ time series.
\item  Step 3: Non-Gaussian modeling. In \cite{scherrer2007}, it was shown that the marginal distribution $f_{\delta}(x)$ of aggregated traffic time series can be described uing Gamma laws $\Gamma_{\alpha_{\delta},\beta_{\delta}}$, which are non-Gaussian distributions for positive random numbers, and which is defined by
        \begin{equation}
		\label{eqn:sketch_gamma}
			\Gamma_{\alpha,\beta} = \frac{1}{\beta \Gamma (\alpha)} (\frac{x}{\beta})^{\alpha - 1}\exp{(-\frac{x}{\beta})}
		\end{equation}
    where $\Gamma(\cdot)$ is the Gamma function. The scale parameter $\beta$ behaves as a multiplicative factor. If $x$ is $\Gamma_{\alpha,\beta}$, then $\gamma x$ is $\Gamma_{\alpha,\gamma\beta}$. The shape parameter $\alpha$ determines the shape of the distribution from a highly asymmetric stretched exponential distribution ($\alpha \rightarrow 0$) to a Gaussian distribution ($\alpha \rightarrow + \infty$). Also, $\Gamma_{\alpha,\gamma\beta}$ distributions are stable under addition. Let $x$ and $x'$ denote two independent $\Gamma_{\alpha,\gamma\beta}$ and $\Gamma_{\alpha',\gamma\beta}$ random variables, then $x + x'$ is $\Gamma_{\alpha + \alpha',\gamma\beta}$-distributed. This is relevant in relation to the aggregation procedure because of the following
        \begin{equation}
		\label{eqn:sketch_agg}
			x_{2\delta}(t) = x_{\delta}(t) + x_{\delta}(t + \delta).
		\end{equation}
    Therefore, if $x_{\delta}$ can be modeled with $\Gamma_{\alpha_{\delta},\beta_{\delta}}$, then $x_{2\delta}$ can be modeled with  $\Gamma_{\alpha_{2\delta},\beta_{2\delta}}$. Independence between $x_{\delta}(t)$ and $x_{\delta}(t + \delta)$ implies that $\alpha_{\delta} = \alpha_0\delta$ and $\beta_{\delta} = \beta_0$. Because of the correlation between $x_{\delta_j}(t)$ and $x_{\delta_j}(t + \delta_j)$, departures of $\alpha_{\delta}$ and $\beta_{\delta}$ from $\alpha_{\delta} = \alpha_0\delta$ and $\beta_{\delta} = \beta_0$ are assured. Thus, the Gamma model combined at various resolutions describes not only the marginal distributions of the aggregated traffic but also its short-time statistical time-dependencies.

    Based on $X_{\delta_j}^{n,m}$, the corresponding set of parameters $(\alpha_{\delta_j}^{n,m}, \beta_{\delta_j}^{n,m})$ are estimated by standard sample moment procedures.
\item  Step 4: Reference. For each $h_n$, the averages and variances are estimated as
        \begin{equation}
		\label{eqn:sketch_avg_var}
			\alpha_{\delta_j}^{m, R} = \langle\alpha_{\delta_j}^{n,m}\rangle\quad \sigma_{m, \alpha, \delta_j}^{2} = \langle\langle\alpha_{\delta_j}^{n,m}\rangle\rangle.
		\end{equation}
where $\langle\cdot\rangle$ and $\langle\langle\cdot\rangle\rangle$ are standard sample mean and variance estimators, computed for each $m$.       \item  Step 5: Statistical distances. Anomalies in $(\alpha_{\delta_j}^{n,m}, \beta_{\delta_j}^{n,m})$ are measured by computing the statistical distance from the reference $\alpha_{\delta_j}^{m, R}$. A number of different statistical distances can be used \cite{basseville1989}. Here, we use Mahalanobis distance, which gives equal weight to all scales, and which is given by
        \begin{equation}
		\label{eqn:sketch_statis_dis}
			(D_{\alpha^{n,m}})^2 = \frac{1}{J}\sum_{j=1}^J \frac{(\alpha_{\delta_j}^{n,m} - \alpha_{\delta_j}^{m,R})^2}{\sigma_{m, \alpha, \delta_j}^{2}}.
		\end{equation}
The anomaly is declared if the following condition holds
        \begin{equation}
		\label{eqn:sketch_anomaly_cond2}
			D_{\alpha^{n,m}} > \lambda
		\end{equation}
where $\lambda$ is the detection threshold to be selected. The use of multi-resolution distance implies the detection is not based on the change in traffic volume but rather on the change in the short-time traffic correlations. Similar procedures are used for parameter $\beta$, but a different distance is used.
\item  Step 6: Anomaly identification by sketch combination. We reverse the hashing procedure to identify the keys associated with the anomalies. When we perform detection in the $m$-th output for the $n$-th hash function, the corresponding attributes $K_i$ are recorded in a detection list $K_i^n$. We combine the $N$ hash functions, and take the intersection of $K_i^n$, which results in a final list of attributes $K_i^o$ that are associated with anomalies. The use of $k$-universal hash function ensures that the probability of collision in hashing diminishes exponentially fast with $N$. We can verify that using $k \geq 4$ and $N = 8$ is adequate for good detection performance.

\end{itemize}

The anomaly detection scheme is evaluated using traffic on a trans-Pacific transit link from 2001 to 2006. The results show this scheme can detect a large number of known and unknown anomalies, whose intensities are low, even down to below one percent.

    \section{Signal-Analysis-Based Approaches}
		\label{section:time_series}
		In this section, we introduce three signal-analysis-based anomaly detection schemes: a statistics-based scheme, a wavelet-based scheme, and a hybrid scheme using filtering and statistical methods. These three schemes are independent of each other. 
\subsection{A Statistics-Based Anomaly Detection Scheme}
In \cite{thottan2003}, statistical analysis was used to detect anomaly in SNMP MIB data at routers. Time-series data are collected at the regular intervals of 5 minutes. Three MIB variables at the IP layer are used: ipIR, ipIDe and ipOR. The variable ipIR, short for In Receives, indicates the total number of datagrams received from all the interfaces of the router. The variable ipIDe, short for In Delivers, represents the number of datagrams correctly delivered to the transport layer with this node being the destination. The variable ipOR, short for Out Requests, indicates the number of datagrams passed from the transport layer to be forwarded by the IP layer. These variables are not independent. The average cross correlation between ipIR and ipIDe is 0.08, that between ipIR and ipOR is 0.05, and that between ipOR and ipIDe is 0.32. In the following, we first describe the how to detect abrupt change in a single MIB variable, then how to combine change detections from multiple MIB variables, and finally how to design the combination operator.
\subsubsection{Abrupt Change Detection} The anomaly detection is performed by detecting abrupt changes in the statistics of three MIB time-series data. Abrupt changes are detected by comparing the residuals obtained from two adjacent windows of data called the learning $L(t)$ and the test $S(t)$ windows. Residuals are obtained by imposing an autoregressive (AR) model on the time-series data. Change detection is done by using a hypothesis test based on the general likelihood ratio (GLR). The GLR $\eta$ for a single variable can be expressed by \cite{leland1994}
        \begin{equation}
		\label{eqn:ts_glr}
		\eta = \frac{\hat{\delta}_L^{-\hat{N}_L}\hat{\delta}_S^{-\hat{N}_S}}{\hat{\delta}_L^{-\hat{N}_L}\hat{\delta}_S^{-\hat{N}_S} + \hat{\delta}_P^{-(\hat{N}_L + \hat{N}_L)}}
		\end{equation}
where $\hat{\delta}_L$ and $\hat{\delta}_S$ are the variances of the residual in the learning and test windows. $\hat{N}_L = N_L - p$, where $p$ is the order of the AR process and $N_L$ is the length of the learning window. Similarly, $\hat{N}_S = N_S - p$, where $N_S$ is the length of the test window. $\hat{\delta}_P$ is the pooled variance of the learning and test windows. The anomaly indicators from individual MIB variables are collected to form an anomaly vector $\phi(t)$, which is a measure of the abrupt changes in the network.

\subsubsection{Combining the Anomaly Vectors} The individual anomaly vectors are combined to generate a measure of anomaly. A linear operator $A$ is used to incorporate the correlations among abrupt changes in the individual MIB variables. In particular, the quadratic functional
        \begin{equation}
		\label{eqn:ts_quad_fn}
		f(\phi(t)) = \phi(t)A\phi(t)
		\end{equation}
is used to create a measure of anomaly, which has a range of [0,1]. The value of 1 indicates maximum anomaly, and 0 no anomaly.

The operator $A$ is an $M\times M$ symmetric matrix, where $M$ is the number of MIB variables and is 3 in our case. $A$ has $M$ real eigenvalues and $M$ orthogonal eigenvectors. A subset of eigenvectors corresponds to the anomalous states in the network. Let $\lambda_L$ and $\lambda_H$ denote the minimum and maximum eigenvalues corresponding to the anomalous states. The anomaly detection problem can be formulated as
        \begin{equation}
		\label{eqn:ts_quad_fn}
		t_a = \text{inf}(t: f(\phi(t)) \geq \lambda_L)
		\end{equation}
where $t_a$ is the earliest time when the functional $f(\phi(t))$ exceeds $\lambda_L$. We declare an anomaly if the above condition is satisfied.

\subsubsection{Design of the Operator $A$} First, we augment the anomaly vector by adding the normal state $\phi_0(t)$ so that all possible network states are included. Thus the network state vector is as follows
        \begin{equation}
		\label{eqn:ts_state_vec}
		\phi(t) = [\phi_0(t), \phi_1(t),...\phi_M(t)].
		\end{equation}
After the augmentation, the operator matrix $A$ becomes $(M + 1)\times (M+1)$-dimensional. Since the normal state is decoupled with the anomalous states, $A$ is a block-diagonal matrix with a $1\times 1$ upper block and $M\times M$ lower block $A_M$. We focus on $A_M$ for the purpose of anomaly detection. The elements of $A_M$ are obtained as follows
        \begin{equation}
		\label{eqn:ts_op_ele}
		A_M(i,j) = |\langle\phi_i(t), \phi_j(t)\rangle| = \frac{1}{T}|\sum_{t=1}^T\phi_i(t)\phi_j(t)|
		\end{equation}
which is the ensemble average of the correlation of the two anomaly vectors estimated over a time interval $T$. For $i = j$, we have
        \begin{equation}
		\label{eqn:ts_op_ele_diag}
		A_M(i,i) = 1 - \sum_{i \neq j}A(i,j).
		\end{equation}

By design, the matrix $A_M$ is symmetric, real, and its elements are non-negative. The eigenvectors $\psi_i$ of $A_M$ are orthonormal. The state vectors $\phi(t)$ can be decomposed as a linear combination of eigenvectors $\psi_i$ as follows
        \begin{equation}
		\label{eqn:ts_decomp_sv}
		\phi(t) =  \sum_{i=1}^Mc_i\psi_i.
		\end{equation}
The above expression provides a decomposition of the anomaly into a number of fault modes, each represented by the so called fault vector $\psi_i$. The operator transformation can be expressed as
        \begin{equation}
		\label{eqn:ts_decomp_trans}
		A_M\phi(t) = \sum_{i=1}^Mc_i\lambda_i\psi_i.
		\end{equation}

The measure of averaged anomaly $E(\lambda)$ is given by
        \begin{equation}
		\label{eqn:ts_avg_anom}
		E(\lambda) = \phi(t)\phi(t) = \sum_{i=1}^Mc_i^2\lambda_i.
		\end{equation}
Let $\lambda_{min}$ and $\lambda_{max}$ denote the smallest and largest eigenvalues of $A$. An anomaly is declared if the following holds
        \begin{equation}
		\label{eqn:ts_anom_decl}
		E(\lambda) > \lambda_{min}.
		\end{equation}

Performance evaluations carried out in \cite{thottan2003} showed the statistics-based method is quite effective in anomaly detection.
\subsection{A Wavelet-Based Anomaly Detection Scheme}
A wavelet-based approach was proposed in \cite{barford2002} for network anomaly detection. The traffic stream, sampled every 5 min, is treated as a generic signal. The wavelet analysis decomposes the traffic into strata. The lower strata contain low-frequency or aggregated information, whereas the higher strata contain high-frequency or fine-grained information.

Wavelet analysis consists of two steps: analysis and synthesis. In the \emph{analysis} step, a hierarchy of strata is extracted from the original data. It works iteratively. Given a signal $x$ with length $N$, the output is two derived signals, each with length $N/2$. Each output signal is obtained by convolving $x$ with a filter $F$. One of  the filters, denoted as $L$, has a smoothing or averaging effect, and generates the low-frequency output $L(x)$. The other filters, $H_1,H_2,...H_m$, perform the discrete differentiation, and generate the high-frequency outputs $H_i(x)$. The process continues with further analysis of $L(x)$, producing shorter signals $L^2(x), H_1L(x), H_2L(x),...H_mL(x)$. In the end, we obtain a family of signals of the form $H_iL^{j-1}$, which are called wavelet coefficients. The index $j$ indicates the number of low-pass filtering performed on the signal. The larger the value of $j$, the lower the signal is in the hierarchy. If the original signal has a sample interval of $\tau$, then $H_iL^{j-1}$ consists of data values that are $2^j\tau$ apart from each other.

In the \emph{synthesis} step, the reverse is performed. At each iteration, inputs are $L^j(x), H_1L^{j-1}(x), H_2L^{j-1}(x),...H_mL^{j-1}(x)$, and the output is $L^{j-1}(x)$. In the end, the original signal $x$ is reconstructed. Sometimes, we can suppress the information we want to ignore by zeroing the values smaller than a threshold in the derived signals. For example, if we wish to view only the fine-grained changes in data, we can apply thresholding to the low-frequency strata.

In selecting filters, we need to consider the balance between \emph{time localization }and \emph{frequency localization}. Time localization can be measured by the length of the filter. Longer filters lead to more blurring in the time domain. High-pass filters need to be short. One measure of frequency localization is the number of \emph{vanishing moments}. We say a filter $H$ has $k$ vanishing moments if $\hat{H}(0), \hat{H}'(0), \hat{H}^{(k-1)}(0) = 0$, where $\hat{H}$ is the Fourier transform of $H$. Every wavelet has at least one vanishing moment. Longer filters have more vanishing moments. Filters with low number of vanishing moments may lead to large wavelet coefficients when no significant event is occurring, resulting in false positives. Thus, we need to avoid such filters. Another measure of frequency localization is the \emph{approximation order} of the system, the definition of which is involved and thus is omitted here. Which measure to use depends on the application on hand. The final factor regarding selecting wavelet systems is {\em artifact freeness}. In some wavelet systems, the reconstructed signal exhibits features that are not part of the original signal but are artifacts of the filters used. Short filters without artifacts are rare.

The wavelet system used in \cite{barford2002} is called PS(4.1)Type II \cite{daubechies2003}, which is a framelet system or a redundant wavelet system where the number of high-pass filters is larger than one, i.e., $m > 1$. In the framelet system, the total number of coefficients exceeds the total length of the original signal. Using framelets, we can construct short filters with good frequency localization.

We use one low-pass filter $L$ and three high-pass filters $H_1, H_2, H_3$. The high-pass filters are all 7-tap, i.e., each having 7 non-zero coefficients. The low-pass filter is 5-tap. The vanishing moments of the high-pass filters are 2, 3, 4, respectively. The approximation order is 4. These filters do not create artifacts.

The \emph{analysis} proceeds as follows. It applies to the Internet traffic data collected every 5 minutes for two months. If the scenario is different, parameters need to be adapted.
\begin{itemize}
\item  The L(ow frequency) part of the signal is obtained by synthesizing all the low-frequency wavelet coefficients from frequency levels 9 and up. The L-part of the signal should capture long-term (several days and up) patterns and anomalies. The L-part of the signal is sparse, and the number of values is 0.4\% of that of the original signal. This part of the signal has a high degree of regularity. Thus, long-term anomalies can be captured reliably.
\item  The M(id frequency) part of the signal is obtained by synthesizing the wavelet coefficients from levels 6, 7, 8. The signal has zero mean. It should capture the daily variations in the data. Its number of values is 3\% of that of the original signal.
\item  The H(igh frequency) part of the signal is obtained by thresholding the wavelet coefficients in the first 5 levels, and setting to zero all the coefficients in levels 6 and up. Most of data in the H-part are small short-term variations, which can be considered as noise. Thresholding helps to get rid of the noise.
\end{itemize}
The \emph{anomaly detection} works as follows.
\begin{itemize}
\item  The H- and M-parts are normalized so that they have unit variance. Then, the local variability of the H- and M-parts is calculated by computing the variance of the data within a moving window, the size of which depends on the duration of anomalies we wish to capture. Let $t_0$ denote the duration of the anomaly, and $t_1$ the size of the window. Ideally, we want $q = t_0/t_1 =1$. If $q$ is too small, the anomaly will be blurred and lost. If $q$ is too large, we will be inundated with "anomalies" that are insignificant.
\item  The local variability of H- and M-part of the signal is combined using a weighted sum. The result is the V(ariable)-part of the signal.
\item  Thresholding is applied to the V-signal. Anomaly detection is performed by measuring the peak height and peak width of the V-signal.
\end{itemize}

The wavelet-based anomaly detection scheme was applied to IP flow and SNMP data collected at the border router at the University of Wisconsin. The results indicate that wavelet-based detection is quite effective at exposing the details of both ambient and anomalous traffic.
\subsection{A Hybrid Anomaly Detection Scheme Based on Filtering and Statistical Methods}
In \cite{soule2005}, a hybrid anomaly detection scheme based on both filtering and statistical methods was proposed. The scheme is composed of two steps. In the first step, Kalman filter is used to filter out the normal traffic. In the second step, statistics-based anomaly detection methods are applied to the residual signal.

\subsubsection{Modeling Normal Traffic} We wish to detect traffic anomalies in the source-destination flows. Traffic volumes of the flows are not directly observable. What are observable (through SNMP) are link traffic statistics. The flow and link traffic volumes are related as follows
        \begin{equation}
		\label{eqn:ts_flow_link}
		Y_t = A_tX_t + N_t
		\end{equation}
where $Y_t$ and $X_t$ are the vectors of link and flow traffic volumes, respectively. The matrix $A_t$ is the routing matrix, whose element $A_{t,i,j}$ is 1 if flow $j$ traverses link $i$, and 0 otherwise. The term $N_t$ represents the measurement noise. All these parameters are defined at time $t$.

A linear state-space model is used to capture the time evolution of the traffic of the flows as follows
        \begin{equation}
		\label{eqn:ts_flow_evol}
		X_{t+1} = C_tX_t + W_t
		\end{equation}
where $C_t$ is the state transition matrix and $W_t$ is the noise process. The diagonal elements of $C_t$ model the correlation in time, and the non-diagonal elements model the correlation between flows.

We assume the measurement and state noises $N_t, W_t$ to be uncorrelated, zero-mean, Gaussian white-noise processes with covariance matrices $R_t$ and $Q_t$:
         \begin{align*}
		E[W_iW_j^T] = \begin{cases} Q_i, & \mbox{if }i = j\\
        0, & \mbox{otherwise}\end{cases}\\
        E[N_iN_j^T] = \begin{cases} R_i, & \mbox{if }i = j\\
        0, & \mbox{otherwise}\end{cases}
        \end{align*}
        \begin{equation}
		\label{eqn:ts_flow_evol}
		E[N_iW_j^T] = 0 \quad \forall i,j.
		\end{equation}
These assumptions may seem restrictive, but Kalman filters are robust to model imprecision and deviation from Gaussianity.

Given observations $Y_1, Y_2,...Y_{t+1}$, the Kalman filter estimates system state $X_{t+1}$ using two steps:
\begin{itemize}
\item  Prediction: Let $\hat{X}_{t|t}$ denote the estimate of the state at time $t$, given the observations up to time $t$. Let $\hat{P}_{t|t}$ denote the variance of $\hat{X}_{t|t}$. Let $\hat{X}_{t+1|t}$ denote the one step predictor, with variance of $\hat{P}_{t+1|t}$. The prediction is performed as follows
    \begin{align}
     \label{eqn:ts_pred}
		\hat{X}_{t+1|t} = C_t\hat{X}_{t|t}\nonumber\\
        \hat{P}_{t+1|t} = C_t\hat{P}_{t|t}C_t^T + Q_t
       \end{align}
\item  Estimation: In this step, the Kalman filter updates the state estimate and its variance by using a combination of the predicted values and the new observation $Y_{t+1}$ as follows
    \begin{align}
     \label{eqn:ts_estim}
    \begin{gathered}
		\hat{X}_{t+1|t+1} = \hat{X}_{t+1|t} + K_{t+1}[Y_{t+1} \notag\\
        - A_{t+1}\hat{X}_{t+1|t}],\\
        \hat{P}_{t+1|t+1} = (I - K_{t+1}A_{t + 1})\hat{P}_{t+1|t}(I - K_{t+1}A_{t+1})^T \\
        + K_{t+1}R_{t+1}K_{t+1}^T.
       \end{gathered}
       \end{align}
    The new state estimate $\hat{X}_{t+1|t+1}$ is computed using the previous prediction $\hat{X}_{t+1|t}$ adjusted by a correction term $K_{t+1}[Y_{t+1} - A_{t+1}\hat{X}_{t+1|t}]$, of which $Y_{t+1} - A_{t+1}\hat{X}_{t+1|t} = Y_{t+1} - \hat{Y}_{t+1}$ is the prediction error and $K_{t+1}$ is the Kalman gain matrix, which is obtained by minimizing the conditional mean-squared error $E[\tilde{x}_{t+1|t+1}^T\tilde{x}_{t+1|t+1}|Y_t]$ and is given by
        \begin{equation}
		\label{eqn:ts_kalman_mtx}
		K_{t+1} = P_{t+1|t}A_{t+1}^T[A_{t}P_{t+1|t}A_{t+1}^T + R_{t+1}]^{-1}.
		\end{equation}
    The above equations together with the initial conditions below
    \begin{align}
     \label{eqn:ts_init}
		\hat{X}_{0|0} = E[X_0]\nonumber\\
        \hat{P}_{0|0} = E[(\hat{X}_{0|0} - X_0)(\hat{X}_{0|0} - X_0)^T]
       \end{align}
    constitute the Kalman filter algorithm.
\end{itemize}

The estimation error, or the innovation, is the difference between the observed value and the predicted value, and is given by
        \begin{equation}
		\label{eqn:ts_kalman_err}
		\epsilon_{t+1} = Y_{t+1} - A_{t+1}\hat{X}_{t+1|t}.
		\end{equation}
The innovation is assumed to be white Gaussian noise with a covariance matrix given by
        \begin{equation}
		\label{eqn:ts_err_cov}
		E[\epsilon_{t+1}\epsilon_{t+1}^T] = A_{t+1}P_{t+1|t}A_{t+1}^T + R_{t+1}.
		\end{equation}

We define the residual $\eta_{t+1}$ as
        \begin{equation}
		\label{eqn:ts_eta}
		\eta_{t+1} = \hat{X}_{t+1|t+1} - \hat{X}_{t+1|t} = K_{t+1}\epsilon_{t+1}
		\end{equation}
The covariance of $\eta_{t+1}$ is given by
        \begin{eqnarray}
		\label{eqn:ts_eta_cov}
		S_{t+1} = E[\eta_{t+1} \eta_{t+1}^T] =\nonumber\\
        K_{t+1}(A_{t+1}P_{t+1|t}A_{t+1}^T + R_{t+1})K_{t+1}^T.
		\end{eqnarray}

The above description of Kalman filtering is in the general setting under non-stationary assumptions. In the following, we assume a stationary setting where the matrices $A, C, Q,$ and $R$ are constant in time and their subscripts are dropped. However, the rest of the methodology can be easily generalized to the non-stationary setting.
\subsubsection{Analyzing Residuals}
There are two sources of errors in the residual process. One is from errors in the underlying traffic model, whereas the other is from the anomalies. Let $Z_t$ denote a general random process that we wish to check for anomalies. Let $\hat{Z}_t$ denote our prediction for the process. Then, we have
        \begin{equation}
		\label{eqn:ts_res_err}
		Z_t = \hat{Z}_t + \zeta_t + \xi_t
		\end{equation}
where $\zeta_t$ and $\xi_t$ are the expected prediction error and the error caused by anomalies, respectively.

We observe the residual $\eta_t$, and a non-zero residual could indicate that an anomaly has occurred. Moreover, the residual $\eta_t$ and the estimation error $\zeta_t$ are correlated Gaussian processes, and we can use one to estimate the other using the following formula
        \begin{equation}
		\label{eqn:ts_eta_zeta}
		\tau_t \doteq \zeta_t + \eta_t \simeq - K_tA_tP_{t|t-1}S_t^{-1}\eta_t.
		\end{equation}
\subsubsection{Detecting Anomalies Using Statistical Test}
We use receiver-operation-characteristics (ROC) curves for assessing the detection performance \cite{egan1975}. ROC curves provide the benefit of presenting the tradeoffs between false positives and false negatives over the full range of operating conditions. In a ROC curve, the x-axis is the false positive rate, and the y-axis is one minus the false negative rate, which is the true positive, i.e., the probability of anomaly. The performance of a detection scheme is considered very good if the ROC curve rises rapidly toward the upper left corner of the graph, which means a large portion of anomalies is detected with low false positive rate. The quantitative measure is the area under the curve. The larger the area, the better the performance.

Given the residual process $\xi_t$, we formulate the anomaly detection as a hypothesis testing problem as follows
        \begin{eqnarray}
		\label{eqn:ts_hypo_test}
		H_0: \xi_t = 0 \quad\text{(no anomaly)}; \nonumber\\
        H_1: \xi_t = \mu \quad\text{(anomalies are detected)}
		\end{eqnarray}
where $\mu > 0$ is some constant. We compare $\xi_t$ to a threshold $T_0$, accept $H_0$ if $\xi_t < T_0$, and accept $H_1$ otherwise.

According to the Neyman-Pearson formulation, the goal is to solve an optimization problem that maximizes the probability of detection while not letting the probability of false alarm exceed a certain value $\alpha$. Let $P_P$ and $P_N$ denote the false positive and false negative probabilities, respectively. The optimal decision threshold is the one that satisfies the likelihood ratio test below
        \begin{equation}
		\label{eqn:ts_ratio_test}
		\frac{P_P}{P_N} \leq T(\alpha)
		\end{equation}
where $T(\alpha)$ is the threshold, which is a function of $\alpha$.

The ROC curve can be obtained analytically only under simple assumptions such as $\xi_t$ is Gaussian.
\subsubsection{Anomaly Detectors}
In the following, we describe four anomaly detectors.
\begin{itemize}
\item  {\em The basic scheme using variance}: We obtain $\tau_t$ using (\ref{eqn:ts_eta_zeta}) and declare an anomaly if the following holds
        \begin{equation}
		\label{eqn:ts_basic_var}
		\tau_t > T_h\sqrt{P_{t+1|t+1}}
		\end{equation}
where $T_h$ is the threshold. This scheme is optimal if $\tau_t$ is Gaussin, and suboptimal otherwise. In this scheme, the test is performed as soon as each new observation is obtained, so it is fast. The downside is that the detection is made independent of past observations, which could lead to high false positive rate.
\item  {\em CUSUM and generalized likelihood ratio test}: This scheme is based on the classical approach for detecting changes in random processes: cumulative summation method (CUSUM) \cite{basseville1993}. In this scheme, a change is declared when the log-likelihood ratio of an observation $y$, defined below, shifts from a negative value to a positive one.
        \begin{equation}
		\label{eqn:ts_cusum_likelihood}
		s_i = \log{\frac{L_1(y)}{L_0(y)}}
		\end{equation}
where $L_1(y)$ and $L_0(y)$ are successive values of the likelihood ratio. Equivalently, the log-likelihood of $N$ observations, defined as $S_{N-1} = \sum_{i=0}^{N-1}s_i$, that was decreasing with $N$, begins to increase after the change, with the minimum $S_j$ providing an estimate of the change point. So, the test for change detection is the following
        \begin{equation}
		\label{eqn:ts_cusum_test}
		S_k - \underset{0\leq j\leq k}{\text{min}} S_j > T_h
		\end{equation}
where $T_h$ is the threshold. The time of the change can be estimated as
        \begin{equation}
		\label{eqn:ts_cusum_time}
		\hat{t}_c = \underset{0\leq j\leq k}{\text{arg min}}\{S_j\}.
		\end{equation}

Although the above CUSUM algorithm has been widely used for anomaly detection, it has a drawback in that it assumes the alternative hypothesis $H_1$ is known a priori, which is not true in many practical scenarios. A solution to the problem is provided by the General Likelihood Ratio Test (GLR) \cite{hawkins2003}. In GLR, the mean of the estimation error over the window $[i, i+1,...j+N-1]$ is estimated as
        \begin{equation}
		\label{eqn:ts_glr_err}
		\hat{\mu} = \frac{1}{j+N-1-i}\sum_{l=i}^{j+N-1}\tau_l.
		\end{equation}
Then, a CUSUM test with $\hat{\mu}$ as the level change value is performed for change detection. It can be proved that GLR is the optimal estimator when the level change value $\mu$ and the variance $\sigma$ are unknown. The downside of GLR is the increased delay caused by the fact that it needs many observations before it can perform the test.
\item  {\em Multi-scale analysis using variance}: The motivation behind multi-scale analysis is that anomalies appear at different time scales, and by monitoring multiple scales we can reduce false positive rate since a change appearing on only one scale will not trigger alarm. Multi-scale analysis can be implemented by a cascade-decomposing of the original signal $\tau_t$ into a low frequency approximation $a_t^L$ and a cascade of details $d_t^i$ as follows
        \begin{equation}
		\label{eqn:ts_multi_decomp}
		\tau_t = a_t^L \sum_{i=1}^{L}d_t^i.
		\end{equation}
    where
        \begin{eqnarray}
		\label{eqn:ts_mult_decomp1}
		d_t^i = \sum_{s}\tau_s 2^{-i}\psi(2^{-i}s -t), i = 1, 2,...L, \nonumber\\
        a_L^i = \sum_{s}\tau_s 2^{-L}\phi(2^{-L}s -t), \nonumber\\
		\end{eqnarray}
    where $\psi(\cdot)$ is a mother wavelet function and $\phi(\cdot)$ is the corresponding scaling function \cite{mallat1999}.

    Anomaly detection is performed using a method similar to the basic scheme using variance. For each level $l$, each time instant $t$ is assigned a bit, where 0 indicates no anomaly and 1 indicate an anomaly is detected. By summing across 0-1 time series, we obtain the number of times that an anomaly was detected across all the details signals. An anomaly is declared if it is detected at a sufficient number of scales. This method introduces delay due to the computation of wavelets.
\item  {\em Multi-scale variance shift}: This method is based on \cite{barford2002}, which is described in the previous section. This method consists of two steps. In the first step, the trend of the signal is removed using a wavelet transform. In the second step, a small window is used to compute the local variance. When the ratio between local variance and the global variance exceeds a threshold, an anomaly is declared. This method is actually a special case of the multi-scale analysis described earlier, where only two scales are analyzed and it detects a variation in the variance rather than the mean. Again, due to the computation of wavelets, this method introduces delay.
\end{itemize}

Performance evaluations performed in \cite{soule2005} showed the GLR performed the best, whereas the wavelet-based method does not perform as well.

    \section{Anomagraphy and Anomaly Extraction}
		\label{section:anomagraphy}
		In this section, we first introduce anomagraphy, which provides a unified frame for synthesizing three major anomaly detection approaches described earlier. Then, we describe anomaly extraction, where the set of flows that caused the anomalies are identified. Anomaly extraction is useful for root cause analysis, attack mitigation, testing anomaly detectors, etc.
\subsection{Network Anomography}
In \cite{zhang2005}, a framework and a class of algorithms were proposed for anomaly inference and detection. The framework subsumes both spatial anomaly detection schemes, such as those based on PCA, and temporal anomaly detection schemes, such as those based on wavelets, statistical analysis, etc. The authors define network anomagraphy as the problem of inferring anomalies from indirect measurement, since anomalies often can not be measured directly. The name anomography comes from combining "anomalous" with "tomography," which is a general approach to inference problems. A dynamic anomography algorithm was introduced, which can effectively track routing and traffic changes. This was the first algorithm that can handle both missing data and routing changes.
\subsubsection{Background}
Network tomography is about making inferences from indirect measurements. Examples include inferring link performance metrics from path performance metrics, and inferring traffic matrices from individual link load measurements. For the latter, in a network that has $m$ links and $n$ source-destination (OD) flows, there is a linear relationship between link loads and the traffic matrices as described below
        \begin{equation}
		\label{eqn:na_tm}
		y = Ax
		\end{equation}
where the $m$-dimensional vector $y$ contains the link measurements, the $n$-dimensional vector $x$ contains OD traffic amounts, and the $m\times n$-dimensional matrix $A$ is the routing matrix, whose element $a_{i,j}$ indicates the fraction of traffic from flow $j$ loaded on link $i$.

To reflect the fact that the traffic of OD flows changes over time, we rewrite the previous formula as
        \begin{equation}
		\label{eqn:na_tm2}
		Y = AX
		\end{equation}
where $Y = [y_1, y_2,...y_t]$ is a matrix collecting link load vectors at a time interval of 5 minutes, and $X = [x_1, x_2,...x_t]$ is a matrix collecting OD flow traffic amounts at the same time intervals. Since we first collect link load matrix and then infer anomalies, it is a batch-processing procedure and is called late-inverse anomography.
\subsubsection{Network Anomography}
A framework of anomography was proposed in \cite{zhang2005}, where the anomalous traffic is extracted by transforming $Y$ and a new set of inference problems are formulated as
        \begin{equation}
		\label{eqn:na_tm2}
		\tilde{Y} = A\tilde{X}
		\end{equation}
where $\tilde{Y}$ and $\tilde{X}$ are the matrices of anomalous traffic and OD flows, respectively.

Two types of transformation are described below.
\begin{itemize}
\item Spatial anomography: A left-multiplying matrix $T$ is used to transform $Y$: $\tilde{Y} = TY$.
\item Temporal anomography: A right-multiplying matrix $T$ is used to transform $Y$: $\tilde{Y} = YT$
\end{itemize}
The details of the transformations are given below.

{\em Spatial Anomography}: In spatial anomography \cite{lakhina2004}, the transformation matrix is given by
        \begin{equation}
		\label{eqn:na_sp_trans}
		T = 1 - P
		\end{equation}
where $P$ is the projection matrix given by (\ref{eqn:pca_u}). The projection maxtrix $P$ projects traffic onto the normal subspace, whereas its complement $1 - P$ projects the traffic onto the anomalous subspace. The above method is called spatial PCA, because it exploits the traffic correlations among different links, i.e., across space. Later, we will introduce temporal PCA that exploits traffic correlations in time.

{\em Temporal Anomography}: In temporal anomography, the transformation matrix can be either explicit or implicit. We consider four examples: AutoRegressive Intergraded Moving Average (ARIMA), Fourier, Wavelet, and PCA.

i) ARIMA: Recall that there are three parameters in the ARIMA model (Section \ref{section:sketch}): the autoregressive parameter ($p$), the number of differencing passes ($d$), and the moving average parameter ($q$). The model can be described by
        \begin{equation}
		\label{eqn:na_arima}
			y_{t,d} - \sum_{i=1}^q {MA}_iy_{t-i, d} = e_t - \sum_{j=1}^p AR_je_{t-j}
		\end{equation}
where $y_{t,d}$ is obtained by differencing the original time series $d$ times, $e_t$ is the forecast error at time $t$, $MA_i$ and $AR_j$ are Moving Average and AutoRegression coefficients.

We can write (\ref{eqn:na_arima}) in matrix form. We consider time series of length $t$. Let $I$ denote the $t\times t$ identity matrix, $\nabla$ denote the $t\times t$ back shift matrix, and $U$ denote the $t\times t$ unit matrix, i.e.,
{\scriptsize
\[ I = \left[ \begin{array}{cccccc}
1&0&0&...&0&0 \\
0&1&0&...&0&0 \\
& & &. . .& &  \\
0&0&0&...&1&0 \\
0&0&0&...&0&1\end{array} \right]
\]
\[ \nabla = \left[ \begin{array}{cccccc}
0 & 1 & 0 & ... & 0 & 0 \\
0 & 0 & 1 & ... & 0 & 0 \\
 &  &  & . . . &  &  \\
0 & 0 & 0 & ... & 0 & 1 \\
0 & 0 & 0 & ... & 0 & 0\end{array} \right]
\]
\[ U = \left[ \begin{array}{cccccc}
1 & 1 & 1 & ... & 1 & 1 \\
1 & 1 & 1 & ... & 1 & 1 \\
 &  &  & . . . &  &  \\
1 & 1 & 1 & ... & 1 & 1 \\
1 & 1 & 1 & ... & 1 & 1  \end{array} \right]
\]
}
The differencing result $y_{t,d}$ is given by
         \begin{align*}
		y_{t,d} = \begin{cases} y(1 - \nabla)^d, & \mbox{if } d \geq 1,\\
        y(I - \frac{U}{t}), & \mbox{if  } d = 0.\end{cases}\\
          \end{align*}
Equation (\ref{eqn:na_arima}) can be expressed in the matrix form as
        \begin{equation}
		\label{eqn:na_arima2}
			y_{t,d} - \sum_{i=1}^q {MA}_i y_{t,d}\nabla^i = e_t - \sum_{j=1}^p AR_je_t \nabla^j,
		\end{equation}
or equivalently,
        \begin{equation}
		\label{eqn:na_arima3}
			e_t = y_{t,d}(1 - \sum_{i=1}^q {MA}_i \nabla^i)(1 - \sum_{j=1}^p AR_j \nabla^j)^{-1}.
		\end{equation}
Thus, the transformation matrix is given by
        \begin{equation}
		\label{eqn:na_arima_T}
			T = (1 - \sum_{i=1}^q {MA}_i \nabla^i)(1 - \sum_{j=1}^p AR_j \nabla^j)^{-1}.
		\end{equation}
In the ARIMA-based anomography, the forecast errors are considered anomalous traffic. In other words, the part of the traffic not captured by the model is considered anomalous.

ii) Fourier Analysis: For discrete-time signals $x_1, x_2,...x_n$, its discrete Fourier transform (DFT) is given by
        \begin{equation}
		\label{eqn:na_ft}
			f_k = \frac{1}{n}\sum_{i=1}^n x_i e^{-\frac{j2\pi i k}{n}},\quad\text{for  }0 \leq k \leq n
		\end{equation}
where $f_k$ is a complex number that represents the signal's projection onto the $k$-th harmonic frequency. For real signal, $f_k$ is symmetric, i.e., $f_k = f_{n-k}$. When $k$ is close to 0 or $n$, $f_k$ corresponds to lower frequency. When $k$ is close to $n/2$, $f_k$ corresponds to higher frequency. The inverse discrete Fourier transform (IDFT) is given by
        \begin{equation}
		\label{eqn:na_ift}
			x_i = \sum_{k=1}^n f_k e^{\frac{j2\pi k i}{n}},\quad\text{for  }0 \leq i \leq n.
		\end{equation}
DFT and IDFT can be efficiently implemented by the fast Fourier transform (FFT) algorithm with a time complexity of $O(n\log n)$.

In general, after performing FFT on the link traffic signal, the low frequency components represent the daily and weekly traffic patterns, and the high frequency components represent sudden changes in traffic pattern. The FFT-based anomography is carried out as follows:
\begin{itemize}
\item  Transform the link traffic $Y$ into the frequency domain, $F = \text{FFT}(Y)$, by applying FFT on each row of $Y$. Recall that each row represents the time series of traffic data on a particular link.
\item  Remove low frequency components, i.e., if $k \leq c$ or $k \geq n - c$, set $F_k = 0$, where $F_k$ is the $k$-th column of $F$ and $c$ is the threshold frequency.
\item  Take the inverse transform back to the time domain, i.e., $\tilde{Y} = \text{IFFT}(F)$, which corresponds to high frequency components that are identified as anomalous link traffic.
\end{itemize}
The DFT, setting columns of $F$ to zero, and IDFT all can be considered as taking linear combinations of the columns of either $Y$ or $F$. Thus, we can write $\tilde{Y} = YT$.

iii) Wavelet Analysis: In Section \ref{section:time_series}, we described a wavelet-based anomaly detection scheme. Such scheme shares the same characteristics as the FFT-based scheme in that they both expose anomalies by filtering low frequency components, but the wavelet-based scheme is superior in situations where the signal contains transients, such as discontinuities and sharp sparks. Specifically, the wavelet-based scheme decomposes the traffic into low-, mid- and high-frequency components, and detects the anomalies by examining the mid- and high-frequency components as described below.
\begin{itemize}
\item  Transform link traffic $Y$ into different frequency components, i.e., $W = \text{WT}(Y)$, by applying wavelet transform on each row of $Y$.
\item  Remove low- and mid-frequency components in $W$ by setting all coefficients at frequency levels higher than $c$ to zero, where $c$ is the threshold frequency level.
\item  Take the inverse transform, i.e., $\tilde{Y} = \text{IWT}(W)$, which results in the high-frequency components that are identified as anomalous traffic.
\end{itemize}
The procedures WT and IWT only involve linear combinations of columns of $Y$ and $W$, respectively. Thus, we can write $\tilde{Y} = YT$.

iv) Temporal PCA: The difference between spatial and temporal PCAs is the following. PCA is performed on $Y$ with spatial PCA, i.e., $\tilde{Y} = (1 - P)Y$, whereas PCA is performed on $Y^T$ with temporal PCA, i.e, $\tilde{Y} = ((1 - P)Y^T)^T = Y (1 - P)^T$, where $P$ is the projection matrix that projects traffic into the normal subspace.

{\em Inference Algorithm}: Given the link anomaly matrix $\tilde{Y}$, we would like to infer the OD traffic flows by solving a series of ill-posed linear inverse problems $\tilde{y_j} = A\tilde{x_j}$. Three algorithms are given below, all of which handle the under-constrained linear problems by seeking a solution that minimizes some versions of vector norm. The $l_p$-norm of an $n$-dimensional vector $x$ is defined as
        \begin{equation}
		\label{eqn:na_lp_norm}
			\|x\|_p = (\sum_{i=1}^n x_i^p)^{1/p}.
		\end{equation}

i) Pseudo-Inverse Solution: The pseudo-inverse solution to the problem $\tilde{y} = A\tilde{x}$ is given by $\tilde{x} = A^{-1}\tilde{x}$, where $A^{-1} \equiv (A^T A)^{-1}A^T$ is the pseudo-inverse of $A$, which exists even when $A$ is not invertible in the normal sense. It is well known that the pseudo-inverse solution minimizes the $l_2$-norm, i.e., the solution solves the following minimization problem
        \begin{equation}
		\label{eqn:na_l2_norm}
			\text{minimize} \quad\|\tilde{x}\|_2 \quad\text{subject to}\quad  \tilde{y} = A\tilde{x}.
		\end{equation}

ii) Sparsity Maximization: Typically, there are few anomalies at any particular time. In other words, $\tilde{x}$ is sparse. Since $l_0$ is equal to the number of nonzero elements in a vector, we can solve the following minimization problem to maximize sparsity
        \begin{equation}
		\label{eqn:na_l0_norm}
			\text{minimize} \quad\|\tilde{x}\|_0 \quad\text{subject to}\quad  \tilde{y} = A\tilde{x}.
		\end{equation}

The $l_0$-norm minimization problem is not convex and is hard to solve. It is well known that an equivalent problem to the $l_0$-norm minimization is the $l_1$-norm minimization, which is given by
        \begin{equation}
		\label{eqn:na_l1_norm}
			\text{minimize} \quad\|\tilde{x}\|_1 \quad\text{subject to}\quad  \tilde{y} = A\tilde{x}.
		\end{equation}

In the presence of the noise, the constraint $\tilde{y} = A\tilde{x}$ does not hold exactly. In such case, we solve the following minimization problem \begin{equation}
		\label{eqn:na_l1_norm2}
			\text{minimize} \quad\lambda\|\tilde{x}\|_1 + \| \tilde{y} - A\tilde{x}\|_1
		\end{equation}
where $\lambda$ controls the tradeoff between the sparsity maximization and the fit to the measurements. It can be shown that the algorithm is not very sensitive to the choice of $\lambda$. The above minimization problem can be cast as an equivalent linear programming problem as below
        \begin{eqnarray}
		\label{eqn:na_l1_norm_lp}
			\text{minimize} \quad \lambda\sum_i u_i + \sum_j v_j \nonumber\\
            \text{subject to}\quad \tilde{y} = A\tilde{x} + z \nonumber \\
            \quad\quad\quad\quad u\geq \tilde{x}, \quad u\geq -\tilde{x} \nonumber\\
            \quad\quad\quad\quad v\geq z, \quad v\geq -z. \nonumber\\
		\end{eqnarray}

iii) Greedy Algorithm: Greedy algorithms such as orthogonal matching pursuit (OMP) can be used to solve the $l_0$-norm minimization problem. For details, the reader is referred to \cite{pati1993}.
\subsubsection{Dynamic Network Anomography}
Previously, we have assumed the routing matrix $A$ is constant. In this section, we consider dynamic routing matrices, i.e., $A_j$ changes over time, which reflects the normal "self-healing" behavior of the network and should be isolated from traffic anomalies. In addition, in practice we have to handle missing data by setting the corresponding rows of $A_j$ to zero.

For the transform-based methods, such as the Fourier, wavelet, and PCA methods, the number of constraints becomes very large as $t$ grows, whereas the same does not grow with the ARIMA model as $t$ grows. Thus, we focus on the ARIMA($p, d, q$) model with $d > 0$, with the case of $d = 0$ omitted for brevity.

Our goal is to seek a solution consistent with the measurements $y_j = A_j x_j$, and an ARIMA model that yields $\tilde{x} = XT$, where $T$ is the same matrix as given by (\ref{eqn:na_arima_T}). Let $L$ denote the back-shift operator, i.e., $L y_t = y_{t-1}$. Let $AR(L)$ denote the AR polynomial as given by
        \begin{equation}
		\label{eqn:na_ar}
			AR(L) = \sum_{i=0}^{d+p}\gamma_i L^i \equiv (1 - \sum_{i=1}^{p}AR_i L^i)(1 - L)^d
		\end{equation}
where $\gamma_i$ is a parameter in the AR model. Let $z_{k-i}=\gamma_i x_{k-i}$. By definition we have
        \begin{equation}
		\label{eqn:na_ar2}
			\sum_{i=0}^{d+p}z_{k-i} = y_{k,d}  - \sum_{i=1}^{p}AR_i y_{k-i,d}.
		\end{equation}
For $d \geq 1$, the ARIMA model can be written as
        \begin{equation}
		\label{eqn:na_ar3}
			\sum_{i=0}^{d+p}z_{k-i} = \tilde{x}_k  - \sum_{i=1}^{p}MA_i \tilde{x}_{k-i}.
		\end{equation}
Let $c_{k-i} = \gamma_i y_{k-i}$, then the measurement equation  $y_j = A_j x_j$ becomes
        \begin{equation}
		\label{eqn:na_ar_meas}
			A_{k-i}z_{k-i} = c_{k-i}.
		\end{equation}

We compute $\tilde{x}_i$ iteratively by solving the following series of minimization problems $P_k$ for $k = 1, 2,...t$:
        \begin{equation}
		\label{eqn:na_ar_min}
			P_k: \quad \text{minimize } \|\tilde{x}_k\|_1\quad \text{subject to  } (\ref{eqn:na_ar3})\text{ and } (\ref{eqn:na_ar_meas}).
		\end{equation}

The techniques described above were evaluated using real traffic data from two large backbone networks \cite{zhang2005}. Among inference algorithms, it was found that algorithms based on the sparsity maximization perform better than the pseudo-inverse algorithm. It was also shown that the dynamic anomagraphy algorithm has few false positives and false negatives, and suffers little performance degradation in the presence of measurement noise, missing data, and routing changes.

\begin{table*}[!ht]
\caption {An Example Frequent Item-Sets}
\begin{center}
\begin{tabular}{ c c c c c c c c c }
\hline
k  & SIP & DIP & SP & DP & \#packets & \#bytes & support & event  \\ \hline
1  & * & * & * & * & 2 & * & 10,407 &                        \\
1  & * & * & * & 25 & * & * & 22,659 &                        \\
2  & Host A & * & * & 80 & 6 & * & 11,800 & HTTP Proxy                        \\
2  & * & * & * & 80 & 6 & * & 35,475 &                        \\
2  & Host B & * & * & 80 & 2 & * & 14,477 & HTTP Proxy                       \\
2  & * & * & * & 80 & 7 & * & 16,653 &                        \\
2  & Host C & * & * & 80 & 2 & * & 15,230 & HTTP Proxy                       \\
2  & * & * & * & 80 & 5 & * & 58,304 &                        \\
3  & * & * & * & 80 & 1 & 48 & 17,212 &                        \\
3  & * & * & * & 80 & 1 & 48 & 11,833 &                        \\
3  & * & * & * & 80 & 1 & 1024 & 23,696 &                        \\
3  & * & * & * & 80 & 1 & 48 & 12,672 &    DoS                     \\
4  & * & Host D & * & 9022 & 1 & 48 & 22,537 &    Backscatter                    \\
5  & * & Host E & 54545 & 7000 & 1 & 46 & 23,799 &    DoS                    \\
5  & * & Host E & 45454 & 7000 & 1 & 46 & 15,627 &    DoS                    \\
\\ \hline
\end{tabular}
\end{center}
\label{table:apriori}
\end{table*}

\subsection{Anomaly Extraction Using Associate Rules}
In \cite{brauckhoff2009}, an anomaly extraction scheme was proposed. In anomaly extraction, first, anomaly detection is performed, and then anomaly extraction is performed, where the set of flows that caused the anomalies are identified. Anomaly extraction is useful for root cause analysis, attack mitigation, testing anomaly detectors, etc. The proposed anomaly extraction scheme first uses histogram-based detectors to identify suspicious flows, and then applies association rule mining to detect anomalous flows. This scheme has the advantage of not requiring the modeling of normal traffic using past data. In the following, we first describe histogram-based anomaly detection, and then describe the application of association rule mining for anomaly extraction.

\subsubsection{Histogram-Based Anomaly Detection}
We consider $n$ histogram-based detectors, each monitoring one of traffic flow features, such as source or destination IP addresses or port numbers. Each histogram has $m$ bins. A technique called {\em histogram cloning} is used. In a traditional histogram, bins group together adjacent values. In the histogram cloning, a hash function is used to randomly place a value to a bin. Each detector uses $k$ histogram clones utilizing independent hash functions. Histogram clones have the benefit of providing multiple views of the network traffic.

During time interval $t$, the Kullback-Leibler (KL) distance between the distribution of the current interval and a reference distribution is computed for each clone. The KL distance measures the similarity of two probability distributions. Given a discrete distribution $q$ and a reference distribution $p$, the KL distance is given by
        \begin{equation}
		\label{eqn:kl_dis}
		D(p||q) = \sum_{i=0}^m p_i\log{(p_i/q_i)}.
		\end{equation}
The reference distribution used here is the distribution from the previous measurement interval.

The first difference of the KL distance time series is normally distributed with zero mean and standard deviation of $\sigma$. We can obtain a robust estimate of the standard deviation $\hat{\sigma}$ from a few training intervals. We raise an alarm if the following holds
        \begin{equation}
		\label{eqn:kl_alarm}
		\Delta_t D(p||q) \geq 3\hat{\sigma}
		\end{equation}
where $\Delta_t$ is the time difference operator and $3\hat{\sigma}$ is the detection threshold. Assume that an anomaly spans multiple time intervals, the anomaly would trigger alarms at its beginning and its end. The above procedure is performed for each feature and for each clone.
\subsubsection{Feature Value Identification}
We use an iterative algorithm to identify the set $B_k$ of histogram bins and the set $V_k$ of feature values that are associated with the anomaly. In each round, we select the bin $i$ with the largest absolute distance max $|p_i - q_i|$ between the current and previous histograms. Then, we simulate the removal of flows in bin $i$ by setting $q_i = p_i$. We test if $\Delta_t D(p||q)$ falls below the threshold. If not, we continue. If yes, we have identified the offending bin and the flow feature value.

There are likely some normal feature values contained in the anomalous bins. In order to reduce false positives, we keep only feature values that have been identified by all histogram clones. A normal feature value appears in all $k$ clones with a small probability of $(1/m)^k$, where $m$ is the number of bins.
\subsubsection{Association Rule Mining}
Association rules describe items that occur frequently together in a data set. Formally, a transaction $T$ consists of a set of $l$ items $T = (e_1, e_2,...e_l)$. The disjoint subsets $X, Y \in T$ define an association rule $X \rightarrow Y$. The support $S$ of an association rule is the number of transactions that contain $X \bigcup Y$.

Discovering association rules in a data set involves two steps: 1) discover the frequent item-sets, which are the item-sets that have a support above a user-specified threshold; and 2) obtain association rules from the frequent item-sets.

The motivation for using association rule mining for anomaly extraction is that anomalous flows have similar flow features, such as IP addresses, port numbers, or flow lengths, since they have a common cause, such as a network failure, a bot engine, or a DoS attack. To apply association rule mining, each transaction $T$ corresponds to a NetFlow record, and the item $e_i$ corresponds to one of the seven $(l = 7)$ flow features: SIP, DIP, SP, DP, protocol, \#packets, \#bytes. For example, the item $e_1$ = (SP = 80) means the source port number is 80, whereas $e_2$ = (DP = 80) means the destination port number is 80. A $k$-item set $X = (e_1, e_2,...e_k)$ is a combination of $k$ items, where $k \leq 7$.

The standard algorithm for discovering frequent item-sets is the Apriori algorithm \cite{agrawal1994}. The algorithm is iterative and makes at most $l$ rounds. In round $i$, it computes the support of the frequent $i$-item-sets. Then the $i$-item-sets whose supports exceed a threshold are selected, which are used as input to construct the $(i + 1)$-item-sets. The algorithm stops when there are no frequent item-set whose support exceeds the threshold. The Apriori algorithm is modified to output only the $i$-item-set that is not a subset of a more specific $(i + 1)$-item-set. This has an effect of significantly reducing the number of item-sets to be processed by a human expert. Let $I$ denote the final set of $k$-item-sets.

An example of using Apriori to extract anomalies is provided in \cite{brauckhoff2009}. A 15-minute trace is used, and the threshold is 10,000 flows. A total of 350,872 flows are flagged as anomalies, the breakdown of which into $k$-item-sets is shown in Table \ref{table:apriori}, where SIP, DIP, SP, and DP stand for source IP address, destination IP address, source port number, and destination port number, respectively. From the table, we can see three anomalies. First, hosts A, B, and C are HTTP proxies that send a large amount of traffic on DP = 80. Second, the traffic on DP = 9022 is a backscatter, since each flow has a different SIP and a random SP. Third, some compromised hosts are launching DDoS attack on host E at DP = 7000. The remaining item-sets are those with common DPs and flow sizes. These item-sets are not anomalous and can be filtered out easily by an administrator.

Evaluation results using the data from a backbone network show the detection cost and false positive rate are significantly reduced by using association rules \cite{brauckhoff2009}.

    \section{Conclusion}
		\label{section:conclusion}
		In this paper, we present three major approaches to non-signature-based network detection: PCA-based, sketch-based, and signal-analysis-based. In addition, we also introduce a framework that subsumes the three approaches and a scheme for network anomaly extraction. We believe network anomaly detection will become more important in the future because of the increasing importance of network security.

\end{document}